\documentclass[epj,nopacs]{svjour}
\usepackage{epsfig,cite}
\usepackage{amsmath}
\usepackage{amssymb}
\usepackage{hyperref}
\usepackage{epstopdf}
\usepackage{graphicx}

\def\beq{\begin{equation}}
\def\eeq{\end{equation}}
\def\ba{\begin{eqnarray}}
\def\ea{\end{eqnarray}}
\def\be{\begin{equation}}
\def\ee{\end{equation}}
\def\K{K{\"a}hler}

\def\ss{\scriptscriptstyle}
\def\gev{{\rm \, Ge\kern-0.125em V}}
\def\tev{{\rm \, Te\kern-0.125em V}}
\def\gyr{{\rm \, G\kern-0.125em yr}}

\def\nl{\hfill\nonumber\\&&}

\def\Toprel#1\over#2{\mathrel{\mathop{#2}\limits^{#1}}}

\def\schi{\widetilde \chi}        

\def\sl{{\widetilde \ell}_{\scriptscriptstyle\rm R}}

\def\stop{\widetilde t}

\def\msf#1#2{m_{\tilde{#1}_{#2}}}

\def\m12{m_{1\!/2}}

\def\mgl{m_{\ss \tilde g}}

\def\mz{m_{\ss Z}}

\def\bea{\begin{eqnarray}}
\def\eea{\end{eqnarray}}

\newcommand{\mchr}[1]{m_{\chi^+_{ #1}}}

\def\mgut{M_{GUT}}
\def\calh{\mathcal{H}}

\def\muS{\mu_{\Sigma}}

\mathchardef\mhyphen="2D

\DeclareMathOperator{\Tr}{Tr}

\begin{document}

\title{Strong moduli stabilization and phenomenology}
\author{ Emilian Dudas\inst{1,2,3} \and Andrei Linde\inst{4} \and
Yann Mambrini\inst{3} \and
  Azar Mustafayev\inst{5,6}
     \and  Keith~A.~Olive\inst{5}}
\institute{Department of Physics, Theory Division, CH-1211, Geneva 23, Switzerland \and
CPhT, Ecole Polytechnique, 91128 Palaiseau, France  \and
Laboratoire de Physique Th\'eorique
Universit\'e Paris-Sud, F-91405 Orsay, France
 \and Stanford Institute of Theoretical Physics and Department of Physics, Stanford University, Stanford, CA 94305, USA
       \and William I.~Fine Theoretical Physics Institute, \\
       School of Physics and Astronomy,
            University of Minnesota, Minneapolis, MN 55455, USA
\and Department of Physics and Astronomy, University of Hawaii, HI 96822, USA}
\date{Received: date / Revised version: date}

\authorrunning{Dudas et al.}
\titlerunning{ Strong moduli stabilization and phenomenology  }

\abstract{
We describe the resulting phenomenology of string theory/supergravity models with strong moduli
stabilization. The KL model with F-term uplifting, is one such example.
Models of this type predict universal scalar masses equal to the gravitino mass.
In contrast,
$A$-terms receive highly suppressed gravity mediated
contributions. Under certain conditions, the same conclusion is valid for gaugino masses, which
like $A$-terms, are then determined by anomalies.
In such models,  we are forced to relatively large gravitino masses (30-1000 TeV).
We compute the low energy spectrum as a function of $m_{3/2}$.
We see that the Higgs masses naturally takes values between 125-130 GeV.
The lower limit is obtained from the requirement of chargino masses greater
than 104 GeV, while the upper limit is determined by the relic density of dark
matter (wino-like).
\vskip 11pt
\begin{center}
CERN-PH-TH/2012-228,
CPHT-RR069.0812,
UMN--TH--3116/12,
FTPI--MINN--12/28,
LPT--Orsay-12-92,
UH-511-1199-12
\end{center}
}

\maketitle
\tableofcontents
\section{Introduction}

One of the goals of this paper is to discuss an interesting interplay between string theory models with moduli stabilization, inflationary cosmology, phenomenological models of supergravity and the mass of the Higgs boson.
Usually string theory is associated with an energy scale which is many orders of magnitude higher than the energies accessible at the LHC. This would make it extremely difficult to test various consequences of string theory.
However, models of moduli stabilization in string theory such as  KKLT~\cite{Kachru:2003aw} allows one to investigate string phenomenology, as well as string cosmology, from a new perspective.
As we will see, that while string theory models with strongly stabilized moduli,
provide natural solutions to several cosmological problems, they lead to a clear separation
in scales in which the effects of string moduli can be tested in low energy experiments.

One of the results found in the simplest versions of the KKLT construction indicates that the mass of the volume modulus, which describes the ``rigidity'' of compactification, is of the same order of magnitude as the gravitino mass \cite{Kallosh:2004yh,Linde:2011ja}. If one  then makes the standard assumption that the gravitino mass is in the TeV range or below it, KKLT constructions bring the scale of supersymmetry breaking in string theory, as well as the masses of some of the the string theory moduli, down to the LHC energy range.
This fact has an interesting phenomenological implication: Supersymmetry breaking in the standard model may be directly affected by details of the KKLT construction. Depending on one's point of view, this may be good news, if one tries to study properties of string theory compactification at LHC, or bad news, if one attempts to make predictions independent of the intricacies of string theory.

More importantly,  this softness of string theory compactification in the simplest versions of the KKLT scenario leads to a specific cosmological problem: vacuum destabilization and decompactification of space  if the Hubble constant during inflation was greater than the gravitino mass  \cite{Kallosh:2004yh}. The requirement $H \leq$~O(1) TeV is extremely restrictive; it would eliminate most (though not all) of the presently existing models of inflation. Moreover, a light volume modulus would lead to a novel version of the cosmological moduli problem, which has plagued supergravity cosmology for more than three decades \cite{Polonyi}. In addition,  one would still need to solve the cosmological gravitino problem, which is another long-standing problem of supergravity cosmology \cite{gravitino,nos}.

A possible solution to the problem of vacuum destabilization was proposed back in 2004, in the same paper where the existence of this problem was expounded \cite{Kallosh:2004yh}. This solution is realized by adding an extra term to the superpotential of the KKLT scenario, as in the well-known racetrack potential. In this construction, the volume modulus mass can be made arbitrarily large, the barrier stabilizing the stringy vacuum can be made arbitrarily high, for any value of the gravitino mass, and the problem of the cosmological vacuum destabilization disappears. To distinguish this model from the original version of the KKLT scenario, we will refer to it as the KL model.

It has often been remarked that the KL model is very fine-tuned. However, a more detailed investigation of this issue in \cite{klor,Linde:2011ja} has demonstrated that the degree of fine-tuning of the parameters of this model is exactly the same as in the standard Polonyi model, or in the original version of the KKLT scenario: It is determined only by the postulated smallness of the gravitino mass.
More specifically, in the KKLT model, the constant term in the superpotential,  $W_{0}$, must be tuned small. In the KL model, we need
approximately the same small number added to a number of O(1) in the superpotential.  Recently, a set of supergravity inflationary models incorporating the KL scenario was proposed, which are very simple and nevertheless are general enough to describe {\it any} set of observational parameters $n_{s}$ and $r$ to be determined by the Planck satellite \cite{Davis:2008fv,Kallosh:2010ug,Kallosh:2010xz,klor}. The KL mechanism of vacuum stabilization can be used also in models of chaotic inflation in string theory as proposed in \cite{Silverstein:2008sg}.

An interesting feature of this class of inflationary models is a controllably small value of the reheating temperature. The gravitino problem may be resolved by a suitably low reheat temperature
or as we will see a large gravitino mass which is imposed by the resulting supersymmetric sparticle spectrum. As for the cosmological moduli problem, supersymmetry breaking is an unavoidable part of the KKLT and KL scenario, which is related to the mechanism of uplifting (see next section). As we shall see, if this mechanism is realized through $F$-term uplifting, no light Polonyi fields are required. This addresses the cosmological moduli problem in the KL scenario, where all moduli can be superheavy.

These advantages of the KL scenario prompted an investigation of its consequences for particle phenomenology  \cite{klor,Linde:2011ja}. The results of this investigation appeared to be much more general than initially expected and apply to the KL model, as well as any other version of the KKLT scenario with strong modulus stabilization.

Because of the strong modulus stabilization, the KL scenario leads to some very specific predictions for supersymmetry breaking and particle phenomenology: \,It describes a certain version of split supersymmetry with anomaly mediation \cite{klor,Linde:2011ja}. Moreover, this prediction is stable with respect to various modifications of the KL model: The same type of supersymmetry breaking and the same pattern of particle masses appears in any version of the KKLT scenario with strong moduli stabilization, which makes the theory cosmologically consistent \cite{Linde:2011ja}. A more precise formulation and explanation of this statement is contained in Section 2, where we give a brief review of the KKLT and KL models.

Moreover, heavy scalars as predicted here are phenomenologically interesting for many reasons. Indeed, it is well known that
heavy squarks can greatly improve the constraints coming from SUSY flavor and CP violating interactions.
In addition, from a theoretical point of view, it is more likely that scalars have
heavier masses than fermions, as fermions can be protected from large radiative corrections whereas scalar particles are generally not protected.

In this paper, we continue an investigation of the low-energy phenomenology of the KL model, as well as all other versions of the KKLT scenario with strong moduli stabilization. After a brief review of moduli stabilization and uplifts in Section 2 and establishing the sources of soft supersymmetry breaking masses in Section 3, we describe the procedure for consistently including radiative electroweak symmetry breaking in this theory.  Because models with strong moduli stabilization
require heavy scalars (as in split supersymmetry \cite{split}) with relatively light gaugino masses (as in models with anomaly mediation \cite{a nom}) there are difficulties in constructing a UV completion for this anomalously split supersymmetric model with the boundary conditions imposed by supergravity. These problems and possible solutions will be discussed in Section 4.  In Section 5, we describe the sparticle mass spectrum in this theory as a function of gravitino mass.  For obvious reasons, we concentrate on the predictions for the Higgs mass in such models. As we will see, to generate a chargino mass of at least 104 GeV (to be consistent with the LEP bound~\cite{LEPsusy}) we need a gravitino mass $m_{3/2} \gtrsim 30$ TeV. At this value, the Higgs mass is $\simeq 125$ GeV, and rises slowly to $\sim 130$ GeV when $m_{3/2} \sim 1000$ TeV. In Section 6, we consider other phenomenological aspects of the model
such as the role of 1 TeV gluinos and their detectabilility at the LHC. We also describe the prospect for dark matter in these models, as well as dark matter detection. Our conclusions are given in Section 7.


\section{Moduli stabilization and uplifting: a brief review}

\subsection{KKLT versus KL}
The KKLT (KL) sector consists of a single chiral field: the modulus $\rho$. We will denote SM fields collectively as $\phi$.
 The scalar potential for uncharged chiral superfields in $\mathcal N=1$ supergravity is  \cite{Fetal}
\ba
V &=& e^{ K}\left({ K}^{a\bar b}D_a W \overline{D_b W} -3| W|^2\right) \ ,
\label{eqn:SUGRApotential}
\ea
where as usual we defined $D_a W = \partial_a W + K_a W$.
We define a \K\ potential with a no-scale \cite{nosc1} structure in the moduli sector and
kinetic terms in the matter sector depending in some unspecified way on the modulus $\rho$.
This can be written as
\be
K = -3 \log (\rho + \bar{\rho}) + h_i^j (\rho, {\bar \rho}) \phi^i \bar{\phi_j} + K(S^i,\bar{S}_i) +  \Delta K(\phi^i,\bar{\phi}_i)
+ {\rm \cdots} \ ,
\label{kahl}
\ee
where $\cdots$ denote terms of higher-order in matter fields $\phi$, irrelevant for our purposes.
In a type IIB string theory setup orientifolded by $\Omega' = \Omega\, I_6 (-1)^{F_L}$, where $I_6$ denotes parity in the six internal dimensions and $(-1)^{F_L}$ is the left-handed fermion number \cite{gkp},  with D7 and D3 branes,  the function $h (\rho, {\bar \rho})$ is a constant if matter fields originate from D7-D7 sector, it is given by $h_i^j (\rho, {\bar \rho}) = \delta_i^j /(\rho + {\bar \rho})$ for fields in the D3-D3 sector and has specific form for fields living at the intersection of various branes.
We will discuss the fields $S_i$ which are associated with $F$-term uplifting below and we specify $\Delta K$ in section 4 in connection with the Giudice-Masiero mechanism \cite{gm}. The important assumption in what follows is that the uplift fields $S_i$ have a separable
\K\ potential, that can be justified, for example, if the uplift fields arise as D7-D7 states.
Indeed, we will be assuming that the $S_i$ are not directly coupled to matter through either
the \K\ potential, the superpotential, or gauge kinetic function.
In each case we will assume the superpotential is separable in the uplift fields
\be
W = W(\rho) + W_F(S^i) + g(\phi^i, \rho) \ ,
\label{super}
\ee
where $W(\rho)$ is either the KKLT or KL superpotential, $W_F$ is the superpotential
associated with uplifting  and $g(\phi^i,\rho)$ is the superpotential for the Standard Model, with $g(0,\rho) = 0$ . The possible $\rho$ dependence of
the matter superpotential $g$ is highly restricted by axionic symmetries and by the origin of matter fields (it is typically an
exponential or a modular form of various weight). Provided that the vev's of matter fields
are very small compared to the Planck scale, the results of the present paper are largely insensitive to the explicit form of the function $h (\rho, {\bar \rho})$ and the $\rho$ dependence of $g(\phi^i, \rho)$.

The superpotential of the KKLT model is
\be
W_\text{KKLT} = W_0 + Ae^{-a\rho} \ .
\label{adssup}
\ee
where $W_0$ and $a>0$ are constants.
In this theory, there is a supersymmetry preserving AdS minimum
found by setting $D_\rho W = 0$. It occurs at ${\rm Im}\, \rho = 0$, and at a certain value $\sigma_{0}$ of the volume modulus $\sigma = {\rm Re}\, \rho$.

After the uplifting to the (nearly Minkowski) dS vacuum state, the gravitino mass becomes
\be
m_{3/2} \approx \sqrt{|V_{\rm AdS}|/3} \approx \frac{aA}{3(2 \sigma_0)^{1/2}}e^{-a \sigma_0} \ .
\label{m32ads}
\ee
Furthermore, after uplifting
\be\label{KKLTDW}
D_\rho W = {3\sqrt 2\over a \sqrt \sigma_{0}}\, m_{3/2} \ .
\ee

The conditions of applicability of the KKLT constructions are $a\sigma_{0} > 1$ and $\sigma_{0} \gg 1$. If one takes $a\sigma_{0}\gg 1$,  the gravitino mass becomes exponentially small. To have $m_{{3/2} } $ in the TeV range in the KKLT scenario, one should take $a\sigma_{0} \sim 30$.

The mass of the volume modulus $\sigma$ in the minimum, as well as the mass of its imaginary (axionic) component,  is given by $
m_{\sigma}  =  2 a\sigma_{0}\, m_{3/2}$ \cite{Linde:2011ja}.
For $a\sigma_{0} \sim 30$, one finds  $m_{\sigma}  \sim 60\, m_{3/2}$. As a result, the mass of the volume modulus is somewhat greater than the gravitino mass, but not by much. This means that the volume stabilization in the KKLT scenario describing light gravitinos is very soft; the mass of the volume modulus in this scenario is many orders of magnitude below the string scale or the Planck scale. It is this softness of the vacuum stabilization that leads to the catastrophic decompactification of extra dimensions during inflation with $H \gtrsim m_{3/2}$ \cite{Kallosh:2004yh,Linde:2011ja}.

The simplest way to avoid this problem is to strongly stabilize the vacuum by making $m_{\sigma}$ greater than $m_{3/2}$ by many orders of magnitude. This was achieved in the KL scenario by using the racetrack superpotential
\be
W_{\rm KL} = W_0 + Ae^{-a\rho}- Be^{-b\rho} \ .
\label{adssuprace}
\ee
In contrast to the KKLT case, the new degree of freedom offered by $Be^{-b\rho}$ allows the new model to have a supersymmetric
Minkowski solution. Indeed,
for the particular choice of $W_0$,
\be\label{w0}
W_0= -A \left({a\,A\over
b\,B}\right)^{a\over b-a} +B \left ({a\,A\over b\,B}\right) ^{b\over b-a} ,
\ee
the potential of the field $\sigma$ has a supersymmetric minimum with
$W_{\rm KL} (\sigma_{0})=0$,  $D_\rho W_{\rm KL}(\sigma_{0}) = 0$,   and $V(\sigma_{0})=0$.

One may add an additional constant $\Delta$ (either positive or negative) to the superpotential (\ref{adssuprace}). This will shift the minimum of the potential down to the AdS minimum with $V_{{\rm AdS}} = -3m_{{3/2}}^2= -{3\Delta^{2} \over 8\sigma_{0}^{3}}$ \cite{klor,Linde:2011ja}, after which one may use uplifting (as in KKLT) to make the cosmological constant as small as $\sim 10^{{-120}}$. Thus one has $m_{{3/2}}^2= {\Delta^{2} \over 8\sigma_{0}^{3}} \ll 1$, which is the only weak-scale fine-tuning required in the KL model. Interestingly, exactly the same level of fine-tuning of the parameter $W_{0}$ is required in the simplest version of the KKLT scenario. This is the standard price for the desire to protect the Higgs mass by the smallness of supersymmetry breaking.

Finally, we turn to uplifting in the theory. Before we make the minuscule addition $\Delta$ to the KL superpotential, supersymmetry is unbroken, the gravitino mass vanishes, but the volume modulus mass is arbitrarily  large, depending on the choice of the parameters $A$, $a$, $B$ and $b$. This mass is virtually unchanged after adding $\Delta$ and uplifting. Thus, one achieves the desired strong vacuum stabilization and removes the cosmological constraint $H \lesssim m_{{3/2}}$. But this strong vacuum stabilization has an interesting implication for the resulting low-energy phenomenology: Just as in the KKLT scenario, $D_\rho W(\rho) = 0$ in the supersymmetric AdS minimum prior to uplifting in the KL model. However, in the simplest version of the KKLT scenario the value of the field $\rho$ does shift slightly during uplifting, and $D_\rho W(\rho)$ becomes (approximately) as large as $W$, as seen in Eq. (\ref{KKLTDW}). In contrast, strong vacuum stabilization keeps moduli practically unchanged during the uplifting. As a result, after the uplifting in the KL scenario, {\it and in any other version of the KKLT scenario with strong vacuum stabilization}, one has $W = \Delta $ and $|D_\rho W(\rho)| \ll |\Delta|, \, m_{3/2}$.

In particular, in the KL model with an uplifting term $\sim \sigma^{-2}$, which appears in the models with uplifting due to anti-$D3$ branes in warped space, one has \cite{Linde:2011ja}
\be\label{KLDW}
D_\rho W =    6 \sqrt{2\sigma_{0}}\,  \frac{m_{3/2}}{m_\sigma} \  m_{3/2} \ .
\ee
On the other hand, one can show that in the F-term uplifting models to be studied below, the uplifting term is proportional to $\sigma^{-3}$, and the result slightly changes,
\be\label{KLDWF}
D_\rho W =    9 \sqrt{2\sigma_{0}}\,  \frac{m_{3/2}}{m_\sigma} \  m_{3/2} \ .
\ee
What is most important for us is that in both cases one has $D_\rho W(\rho) \ll W$, $D_\rho W(\rho) \ll m_{3/2}$, under the condition of strong vacuum stabilization $m_{\sigma} \gg m_{3/2}$, which ensures vacuum stability during high energy inflation.
Thus,  independently of the particular choice of the stabilizing superpotential $W(\rho)$, one can simply take  $D_\rho W(\rho) = 0$ at the minimum of the potential, before and after the uplifting. And this means, as one can easily check, that in all models of such type one has, at the minimum of the potential,
\be\label{Wmin}
|W| =  |\Delta|  = (2 \sigma_{0})^{{3/2}}\, m_{{3/2}}
\ee
and
\be\label{Wprime}
W_\rho =  {3\Delta\over 2\sigma_{0}} \ .
\ee

\subsection{F-term uplifting examples}

In the discussion above, we assumed uplifting as an effect arising solely from
string theory, for example, through the energy of anti-$D3$ branes placed in a highly warped throat.
In the KL model when coupled to matter, the suppressed $F$-term given in equation (\ref{KLDW}) leads to
extremely small values for gaugino masses and tri-linear supersymmetry breaking $A$-terms
which are proportional to $D_\rho W$. In \cite{Linde:2011ja},
the coupling of matter to the uplifting term
was neglected and it was assumed that soft scalar masses remained equal to the gravitino mass.
However, as can be seen from the analysis in Ref. \cite{Choi:2005ge}, there is a cancelation
which leaves only a tiny scalar mass also of order $m_{3/2}^2/M_P$.
The resulting spectrum would then be dominated purely by anomaly mediation,
and suffer from known phenomenological problems \cite{badanom,bbm}.

However, it is in fact relatively simple to recover the result given in \cite{Linde:2011ja},
by using $F$-term uplifting \cite{scrucca,dpp,Kallosh:2006dv,abe} instead of antibranes.
In fact this possibility is quite generic and can be done in various ways, as we now describe in two explicit examples.
The main idea is to use a SUSY breaking sector for uplifting, preferably with a dynamical scale leading to a small mass parameter $M << 1$,
which breaks SUSY in the rigid limit $M_P \rightarrow \infty$. All masses in this sector will be determined, at the tree and one-loop level,
by the dynamical scale and are much larger than the gravitino mass. Whereas in the rigid limit the uplift sector is decoupled from
the KL sector, supergravity interactions couple the two sectors. However, in strong moduli stabilization models like KL, provided
the KL modulus mass and masses of uplifted fields are much larger than gravitino mass, supergravity interactions only change
the original KL and uplift sector minima in a very tiny way. As a result, the vacuum structure is essentially unchanged and the modulus sector contribution to SUSY breaking is completely negligible, as will be seen in what follows.

This implies that in the models with strong stabilization, there is a certain decoupling of string theory moduli from the standard model phenomenology. While this could seem almost obvious on general grounds, it is not the case in the simplest versions of the KKLT scenario. Optimistically, this means that investigation of the low energy phenomenology may provide us with a possibility to test various mechanisms of moduli stabilization in string theory.

Furthermore, in what follows we will be considering models of strong stabilization for {\em all}
moduli, including the $F$-term uplifting fields $S_i$.  Just as strong stabilization
of the volume modulus $\rho$ was advantageous for cosmology and inflation, strongly stabilized
uplifting fields, provides a simple mechanism to avoid cosmological problems associated with
these moduli. While we do not enter into the details of the cosmology of the these moduli,
we take the premise that all moduli are strongly stabilized.

\subsubsection{\bf F-term uplifting with a non-minimal Polonyi field}

We begin with a very simple example based on a non-minimal version of
the Polonyi model, known as O'KKLT \cite{dine,Kallosh:2006dv}.
The O'KKLT model for $F$-term uplifting is realized with the following
definitions of  $K(S,\bar{S})$ and $W_F(S)$ used in eqs. (\ref{kahl}) and (\ref{super})
for a single Polonyi-like field $S$.
We take
\be\label{kalpol}
 K(S,\bar{S}) = S \bar{S} - \frac{(S \bar{S})^{2}}{\Lambda^2} \ ,
 \ee
 where we assume that $\Lambda \ll 1$ (in Planck units). As we will see, the second term
 in (\ref{kalpol}) provides strong stabilization for the field $S$.
 For the superpotential, we can take simply,
 \be\label{superPol}
 W_F(S) =  M^2 S \ ,
 \ee
as in the Polonyi model, but without an additional constant which is necessary for the fine-tuning of the vanishingly small value of the cosmological constant. This constant is already provided by the KKLT/KL superpotential.

In the original O'KKLT model, it was assumed that the term $-\frac{(S \bar{S})^{2}}{\Lambda^2}$  appears after integrating out some heavy degrees of freedom in the O'Raifeartaigh model. A consistency of this assumption required careful investigation  \cite{Kallosh:2006dv}. However, assuming that this interpretation of the term $-\frac{(S \bar{S})^{2}}{\Lambda^2}$ is available, one can simply consider this term as a part of a modified Polonyi model (\ref{kalpol}), (\ref{superPol}) without further discussion of its origin \cite{dine,Kallosh:2006dv}.

The simplest way to understand the main idea of this scenario is to consider it in the context of the KL model, or any other strongly stabilized model of that type. In this case, the position of the AdS minimum of the potential is strongly fixed. Therefore it is not affected by adding the Polonyi field to the theory. In order to find the value of the Polonyi field and its superpotential, it is sufficient to calculate the values of the superpotential $W(\rho)$ of the KL model and its derivative $W_\rho(\rho)$ at the minimum of the KL potential ignoring the Polonyi fields. The results of these calculations are given in (\ref{Wmin}) and (\ref{Wprime}). These results are then used  in the calculation of the F-term potential of the field $S$.

Alternatively, one may wish to abandon any connection to string theory and
simply consider the supersymmetry breaking sector of the Polonyi field $S$ with
strong stabilization provided by the
\K\ potential and superpotential given by Eqs. (\ref{kalpol}) and (superpol).
For the superpotential, however, we must add back the constant term.
For small $\Lambda$, we have strong stabilization and
the mass of $S$ can be large, as discussed below and its
expectation value close to 0.  So long as we continue to assume that the gauge kinetic
function does not linearly depend on $S$, the phenomenological results discussed below
will be unchanged.

These  calculations, for strongly stabilized theories, show that the field $S$ uplifts the AdS minimum to the nearly Minkowski minimum for
\be
M^{4} = 3 \Delta^{2}  = 24\sigma_{0}^{3} m_{{3/2}}^2\ ,
\ee
which determines the choice of the parameter $M$ in (\ref{superPol}).

The field $S$ at the minimum of its potential is real, and its value is given by
\be
\langle S \rangle = {\sqrt 3 \Lambda^{2}\over 6} \ .
\ee
The  mass squared of the field $S$ in both directions (real and imaginary) is given by
\be
m_{S}^{2} = {3\Delta^{2}\over 2 \sigma_{0}^{3}\Lambda^{2}} = {12 m_{{3/2}}^2\over \Lambda^{2}} \gg m^{2}_{3/2} \ ,
\ee
so it too is strongly stabilized. This is quite important. Indeed, the cosmological moduli problem appears because in the minimal Polonyi field model, the mass of the Polonyi field $S$ is of the same order as the gravitino mass, which was supposed to be in the range of $1$ TeV or below. In our model, $m_{S}^{2} \gg m^{2}_{3/2}$ and the field is constrained to lie
close to its minimum near $S = 0$ (for small $\Lambda$).
Moreover, as we will soon see, in the models of this class one typically has $m_{3/2} \gg 1$ TeV. Therefore for sufficiently large $m_{3/2}$ and sufficiently small $\Lambda\ll 1$, the cosmological moduli and gravitino problems will disappear.

Strong stabilization of the field $S$  is important in another respect as well. Since the field $S$ is strongly stabilized, we can repeat the same procedure that we used before,  and calculate the soft breaking terms in the standard model. The only additional parameters that we need for these calculations are the values of $W_{F}$ and $W_{F}'$ at the minimum of the potential for the field $S$, ignoring the standard model fields:
 \be
 W_{F} =    {|\Delta|\Lambda^{2}\over 2}
 \ee
and
\be
\partial_S W_{F} = \sqrt 3\, |\Delta| \ .
\ee
As a result,
\ba
W & = & W(\rho) + W_F = \Delta + \frac{\Delta \Lambda^2}{2} \simeq \Delta\ , \nonumber \\
D_S W & = & \partial_S W_{F} + K_S \ (W(\rho) + W_F) \nonumber \\
& = & \sqrt{3} |\Delta| + \frac{\sqrt{3}}{6} \Lambda^2 (1+ \frac{1}{2}\Lambda^2) |\Delta |
\simeq  \sqrt{3} |\Delta| .
\label{DSW}
\ea

Finally, we should note that in the context of this model, one can attribute the supersymmetry breaking parameter $\Delta$ either to the KL model, as we did earlier, or to the Polonyi model, by adding it to the superpotential (\ref{superPol}). Alternatively, one may add $\Delta_{\rm KL}$ to the KL superpotential (\ref{w0}) and $\Delta_{\rm P}$ to the Polonyi superpotential (\ref{superPol}). The final results and the standard model phenomenology will depend only on the sum of these two parameters, $\Delta = \Delta_{\rm KL}+ \Delta_{\rm P}$. This is another way to see that the degree of fine-tuning required in the KL model is exactly the same as in the Polonyi model.
\subsubsection{\bf F-term uplift with a dynamical ISS sector}

As a second example, we display here another $F$-term uplifting \cite{dpp,llmnr} via the ISS mechanism
\cite{iss}, which leads to a qualitatively similar result to the one in the previous section. The difference is that
in the present example the corresponding mass scale $M$ has a dynamical origin, that naturally explain its smallness.
The model is defined by
\begin{eqnarray}
&& W \ = \ W_{\rm KL} (\rho) \ + \ W_F  (\chi^i) \ , \nonumber \\
&& K \ = \ - 3 \ \ln (\rho + {\bar \rho}) \ + \ |q|^2 \ + \ |{\tilde q}|^2 \ + \  |S|^2 \ .
\label{iss1}
\end{eqnarray}
In (\ref{iss1}),  $\chi^i$ denotes collectively the fields $q_i^a$, ${\tilde q}_a^{\bar j}$,
$S_{\bar j}^i$ of the ISS model \cite{iss}, where $i,{\bar j} = 1 \cdots N_f$ are flavor indices and $a,b = 1 \cdots N$ are color indices. Moreover, in (\ref{iss1})
\begin{eqnarray}
&& W_F (\chi^i) \ = \ h \ Tr \ {\tilde q} \ S \ q \ - \ h \ M^2 \ Tr S \ , \label{iss2}
\end{eqnarray}
and $W_{\rm KL}$ is given in Eq. (\ref{adssuprace}).
As explained in \cite{iss}, the sector  $q_i^a$, ${\tilde q}_a^{\bar j}$ has a perturbative description in the free magnetic range $N_f > 3 N$. SUSY is broken in (\ref{iss2}) by the ``rank condition", i.e. F-terms of meson fields
$(F_{S})_i^j = h {\tilde q}_a^j q_i^a - M^2 \delta_i^j$ cannot be set simultaneously to zero.

As is transparent in (\ref{iss1}), the KL and the ISS sectors are
only coupled through gravitational interactions. In the type II orientifold setup, if the ISS gauge group comes from D3 branes,
the dynamical scale in the electric theory and therefore also the mass parameter $M$ in the magnetic theory
superpotential (\ref{iss2}) depend on the dilaton $S$, which we assume is already stabilized by NS-NS and RR
three-form fluxes \cite{gkp}. As in the O'KKLT model, this decoupling between the uplift field(s) and modulus $\rho$
is instrumental in getting the uplift of the vacuum energy.

At the global supersymmetry level and before gauging the color
symmetry, the ISS model has a global symmetry $G = SU(N) \times
SU(N_f)_L \times SU(N_f)_R \times U(1)_B \times U(1)' \times U(1)_R$,
broken explicitly to $ SU(N) \times SU(N_f)_V \times U(1)_B \times
U(1)_R$ by the mass parameter $M$. In the supergravity embedding
(\ref{iss2}), the R-symmetry $U(1)_R$ is explicitly broken.
We consider here only the ungauged theory for simplicity, in which
the $SU(N)$ is part of the global symmetry group. For the effects of the gauging, see e.g. \cite{dpp} in the related context of the
KKLT uplift.  At the global supersymmetry level, the metastable ISS vacuum is
\begin{equation}
S_0 \ = \ 0 \ , \qquad  q_0 \ = \ {\tilde q}_0^T \ =
 \left(
\begin{array}{c}
M I_N \\
0
\end{array}
\right)
 \ , \label{iss3}
\end{equation}
where $I_N$ is the $N \times N$ identity matrix and $M \ll
\Lambda_m$, where $\Lambda_m \le M_P$ denotes the mass scale
associated with the Landau pole for the gauge coupling in the
magnetic theory. The first question to address is the vacuum
structure of the model. In order to answer this question, we start
from the supergravity scalar potential (\ref{eqn:SUGRApotential}).
By using (\ref{iss1})-(\ref{iss2}), we find
\begin{eqnarray}
V \ =&& \ {e^{{\bar \chi}_{\bar i} \chi^i} \over (\rho + {\bar \rho})^3} \
\left\{  {(\rho + {\bar \rho})^2 \over 3} \big\vert D_{\rho} W \big\vert^2
\nonumber
\right.
\\
&&
\left.
+
\sum_i \big\vert \partial_i W \ + \ {\bar \chi }_{\bar i} W \big\vert^2 \ - \ 3 |W|^2  \right\} \ . \label{iss4}
\end{eqnarray}

Since $M \ll M_P, $ the vev's in the ISS model are well below the Planck scale. Then an illuminating way of rewriting
the scalar
potential (\ref{iss4}) is to expand it in powers of the fields $\chi^i/M_P$, in which case it reads\footnote{In most
of the formulae of this letter, $M_P=1$. In some formulae, however, we keep explicitly $M_P$.}

\begin{eqnarray}
& V& \ = \ {1 \over  (\rho + {\bar \rho})^3} \ V_{ISS} (\chi^i, {\bar \chi}_{\bar i}) \ + \ V_{KL} (\rho,{\bar \rho}) \ + \
{ {\bar \chi}_{\bar i}  \chi^i \over M_P^2} \ V_1  (\rho,{\bar \rho}) \ \nonumber \\
& +& \ {1 \over M_P^3} \ \left[ \ W_{\rm ISS} (\chi^i) \ V_2 (\rho,{\bar \rho}) + \chi^i \partial_i W_{\rm ISS} \
V_3 (\rho,{\bar \rho}) \ + \ h.c. \right]  \nonumber \\
&+& \ \cdots \ , \label{iss5}
\end{eqnarray}
\noindent
where by comparing (\ref{iss5}) with (\ref{iss4}) we can check that
$V_1 \sim m_{3/2}^2 M_P^2$,  and $V_2, V_3 \sim m_{3/2} M_P^3$. The contribution to the vacuum energy from the ISS sector, in the global limit, is $ \langle V_{\rm ISS} \rangle \ = \ (N_f-N) \ h^2  \ M^4 $.
Since we are interested in $30-1000$ TeV scale gravitino masses, it is clear that the first two terms in the rhs of
(\ref{iss5}), $V_{ISS}$ and $V_{KL}$, are the leading terms. Consequently, there should be a vacuum very close to an uplift KL vacuum $\langle \rho \rangle = {\rho}_0$
and the ISS vacuum $\langle \chi^i \rangle = \chi^i_0 $.
The cosmological constant at the lowest order is  given by
\begin{equation}
\Lambda \ = \ V_{KL} (\rho_0, {\bar \rho}_0) \ + \ {(N_f-N) h^2 M^4 \over (\rho_0 + {\bar \rho}_0)^3} \ , \label{iss7}
\end{equation}

\noindent
which shows that the ISS sector plays indeed the role of un uplifting sector of the KL model. In the zeroth order approximation
and in the large volume limit $\sigma_0  \gg 1 $, we find that the condition of zero cosmological constant
$\Lambda = 0$ implies roughly
\begin{equation}
3 \ |W|^2 \ \sim \ h^2 \ (N_f-N) \ M^4 \ . \label{iss8}
\end{equation}
If we want to have a gravitino mass in the $30-1000$ TeV range, we need small
values of $M \sim (10^{-5} - 10^{-6}) M_P$. Since $M$ in the ISS model
has a dynamical origin, this is natural. Moreover, the
metastable ISS vacuum has a significantly large lifetime
exactly in this limit. Therefore,  the claimed value of the gravitino mass is
natural in our model and compatible with the uplift of the
cosmological constant.

\noindent
In the rigid $M_P \rightarrow \infty$ limit, the ISS fields have masses of order
\begin{eqnarray}
&& {\rm tree-level} \qquad m_0 \sim |h M | \ , \nonumber \\
&&  {\rm one-loop} \qquad m_1 \sim \frac{|h^2 M|}{4 \pi} \ . \label{iss9}
\end{eqnarray}

\noindent
Of course the goldstone bosons of the broken global symmetries are massless for the time being. It is easy to remove these massless states by breaking the global symmetry from the very beginning by having several mass parameters,
$M^2 Tr S \rightarrow \sum_i M_i^2 S_{i}^i$.
Notice that supergravity corrections give tree-level masses to the
pseudo-moduli fields of the ISS model. As explained in more general
terms in \cite{iss}, these corrections are subleading with respect
to masses arising from the one-loop Coleman-Weinberg effective
potential in the global supersymmetric limit. This can be explicitly
checked starting from the supergravity scalar potential (\ref{iss4})
and expanding in small fluctuations around the vacuum (\ref{iss3})
to the quadratic order.

Similarly to the previous O'KKLT example, there is no moduli problem in the present setup: both the $\rho $ modulus and the ISS fields are much heavier than the gravitino mass.

\section{Soft masses for matter fields}

While the particular form of the KL superpotential was instrumental in our analysis, the relation $D_\rho W(\rho) \ll W$, which we use in this section for
computing soft terms for matter fields has a much more general validity. It follows directly from our requirement of strong vacuum stabilization, which solves the problem of decompactification during inflation with $H \gtrsim m_{3/2}$, as well as the cosmological moduli problem.

An additional assumption we will make in what follows is that there is no direct coupling in the \K\ potential, superpotential and gauge kinetic function between matter fields and the uplift fields. The absence of linear couplings to the SUSY breaking uplift fields in the gauge kinetic function and superpotential for matter fields can be argued at various levels: \\
- At the level of symmetries in the second uplift example based on ISS model, the meson
fields $S_i^j$ there transform under chiral symmetries $SU(N_f)_L \times SU(N_f)_R$, broken only by mass terms. It is expected that couplings to MSSM fields respect chiral symmetries of the uplift sector, therefore linear couplings to $S$ should be absent.\\
- The uplift Polonyi or ISS sector does break SUSY in the rigid limit in the absence of additional couplings to matter and moduli fields.
When these additional couplings are present, supersymmetry tends to be restored, especially for those couplings which break the R-symmetry
of the uplift sector. It is possible that  the vacuum we are discussing will still be a local miminum with a very long lifetime, however the absence of new couplings helps in avoiding new supersymmetric minima (this argument, however, 
does not pertain to the gauge kinetic function). \\
- From a string theory viewpoint,  the  linear terms in superpotentials present in both of our examples do not arise at tree-level in string
perturbation theory. They can arise nonperturbatively by
D-brane instanton effects. In this case S is actually a field charged under an ``anomalous" $U(1)_X$. This $U(1)_X$ is
broken close to the string scale by field-dependent Fayet-Iliopoulos
terms, depending on some modulus field called $T$ in what follows.
The axionic field in the $T$ multiplet  is shifted nonlinearly under
$U(1)_X$,  $T \rightarrow T + i\, \delta_{GS}\, \alpha$, where $\alpha$ is the gauge transformation parameter,
and is eaten up by the $U(1)_X$ gauge field. At the perturbative level,
couplings of $S$ are very restricted by the $U(1)_X$ symmetry. The gauge
kinetic function $h_A$ must clearly be invariant, and therefore $S$ cannot appear
there perturbatively.   Instantonic effects are proportional to the D-instanton action
$S_{\rm inst} = e^{- 2 \pi T}$, which has a specific $U(1)_X$ charge.
Linear terms in $S$ can arise nonperturbatively in $h_A$ and $W$ via
the gauge invariant combination  $ e^{- 2 a T } S$, where the $U(1)_X$ charge
of  $e^{- 2 a T }$ compensates that of $S$ \cite{linear}. In our uplift examples, it would mean that $S$ couplings in $h_A$ and
the $S$-dependence of Yukawas are suppressed by the mass parameter $M^2 \sim e^{- 2 a T }$. For example $h_A = h_A^0 (1 + \beta_A e^{-2aT} S)$ or
$y_{ijk} = y_{ijk}^0 (1 +  c_{ijk} e^{- 2 a T } S)$. 
However, the ``anomalous" symmetry does not forbid couplings in the \K\ potential, which we have argued against earlier.

Under the assumptions above we now show that strong moduli stabilization with any F-term uplifts leads to small $A$-terms which are dominated by anomaly contributions.
As we will see, this fact alone forces one to
large scalar masses and hence a large gravitino mass.
This is acceptable if the symmetry preventing a linear coupling
of $S$ to matter is operative, and hence we are restricted to small
gaugino masses also dominated by anomaly contributions.
Couplings
in the \K\ potential of the type $S^{\dagger} S \phi^{\dagger} \phi$, on the other hand, are invariant under all symmetries and, if present,  they can change scalar masses in what follows. We will comment on their possible effects below.

Soft terms for matter fields generated in supergravity in the limit $M_P \to \infty$ with fixed gravitino mass $m_{3/2}$ \cite{bfs}
have a nice geometrical structure. For F-term SUSY breaking, they are given by \cite{kl}
\ba && m^2_{i \bar{j}} = m_{3/2}^2~( G_{i \bar{j}} -
R_{i\bar{j}\alpha\bar{\beta}} G^\alpha G^{\bar{\beta}} ~) \ , \nonumber \\
&& (B \ \mu)_{ij} = m_{3/2}^2~ (2  \nabla_i G_j + G^\alpha \nabla_i
\nabla_j G_\alpha ) \ , \nonumber \\
&& (A \ y)_{ijk}= m_{3/2}^2 \left( 3 \nabla_i \nabla_j G_k + G^\alpha
\nabla_i \nabla_j \nabla_k G_\alpha  \right) \ , \nonumber \\
&& \mu_{ij} = m_{3/2} \ \nabla_i G_j  \ , \nonumber \\
&& m_{1/2}^a = { 1 \over 2} ( Re \ h_A)^{-1} m_{3/2} \ \partial_{\alpha}
h_{A} \ G^\alpha \ , \label{dis1} \ea
where $G = K + \ln |W|^2$,
$y_{ijk}$ are Yukawa couplings, $h_A$ are the gauge kinetic functions and
$\nabla_i$ denotes K\"ahler covariant derivatives
\beq
\nabla_i G_j = \partial_i G_j - \Gamma_{ij}^k G_k \, ,
\eeq
where $\Gamma_{ij}^k = G^{k \bar l} \partial_i G_{j \bar l}$
 is the K\"ahler connection. Greek indices $\alpha,\beta$ in (\ref{dis1}) refer to SUSY breaking fields $S$ and $\rho$,
 latin indices refer to matter fields, whereas
$R_{i\bar{j}\alpha\bar{\beta}}$ is the Riemann tensor of the K\"ahler space spanned by chiral (super)fields.
In our models with strong moduli stabilization and decoupling between uplift fields and matter fields, the curvature terms in
the scalar masses of matter fields are negligible and we find to great accuracy
\be
m_0^2 = m_{3/2}^2 \ , \label{m0}
\ee
where the gravitino mass is given by
\beq
m_{3/2}^2 = e^G = \frac{1}{8\sigma_0^3} |W(\rho)+ W_F(S)|^2\, ,
\eeq
and fixes the universal mass scale for scalars. For the O'KKLT model described above, so long as $\Lambda^2 \ll 1$, the dominant contribution
to the gravitino mass comes from $W(\rho) = \Delta$ at the minimum (see eq. (\ref{Wmin})). The trilinear terms are given by
\beq
(A y)_{ijk}   =  e^K K^{\alpha \bar{\beta}} \overline{ D_{\beta } W} ( K_{\alpha} + \nabla_{\alpha}) W_{ijk} \ ,
\label{aterm}
\eeq
where $W_{ijk} = \partial_i \partial_j \partial_k g$, where $g(\phi^i,\rho)$ is the superpotential for matter fields (\ref{super}).
In our case, more explicitly they equal
\be
(A y)_{ijk}= e^K \left[ K^{\rho \bar{\rho}} \overline{ D_{\rho } W} ( K_{\rho} + \nabla_{\rho})
+ K^{S \bar{S}} \overline{ D_{S } W} K_S  \right] W_{ijk}    \, ,
\ee
where we used, according to  the arguments given above, our hypothesis that Yukawas depend very weakly on $S$. For bilinears B-terms, keeping also Giudice-Masiero like terms, we find
\ba
&& (B \mu)_{ij} = e^K K^{\alpha \bar{\beta}} \overline{ D_{\beta } W} ( K_{\alpha} + \nabla_{\alpha}) W_{ij} - m_{3/2} e^{K/2} W_{ij} +   \nonumber \\
&& m_{3/2}^2 (2 + G^{\alpha}
\nabla_{\alpha}) K_{ij}  -m_{3/2}^2  \Gamma_{ij}^{\alpha} (2 G_\alpha + G^\beta G_{\alpha \beta}) \ . \label{bmu}
\ea
Notice that in our case, since $D_{\rho} W$ and $K_S \sim \bar S $ are very small, we find negligibly small A-terms.
More precisely,  we find that the dominant contribution to $A_0$ is given by  ${\bar S} {\bar D}_{\bar S}  \bar{W}$
so that at the tree-level one finds that the $A$-terms are given by
\be
A_0 \simeq -\frac{1}{(2\sigma_0)^{3/2}} \frac{|\Delta| \Lambda^2}{2} = \frac{1}{2}m_{3/2} \Lambda^2
\ee
and are extremely small if $\Lambda \ll 1$.
This expression for $A_0$ is valid so long as $m_{3/2}/m_\sigma \ll \Lambda^2\ll1$. For $\Lambda^2\ll m_{3/2}/m_\sigma \ll 1$ the parameter $A_{0}$ is proportional to $m_{3/2}^{2}/m_\sigma$, so in both cases $A_{0} \ll m_{3/2}$.
Thus we are driven to small values of $A_0$ as a direct consequence of strong stabilization.
On the other hand, the $\mu$ and $B\mu$ parameters in the Higgs sector  are given by
\ba
&& \mu = m_{3/2} G_{12} = e^{K/2} W_{12} + m_{3/2} K_{12} = \mu_0 +  m_{3/2} K_{12}
\ , \nonumber \\
&& B \mu = (A_0 - m_{3/2}) \mu_0 + 2 m_{3/2}^2 K_{12} \ . \label{bmu2}
\ea
where $W_{12} = \partial_{H_1} \partial_{H_2} W$,  $K_{12} =  \partial_{H_1} \partial_{H_2} K $,
and  $\mu_0 = e^{K/2} W_{12} $ is the $\mu$-term in the absence of Giudice-Masiero terms.
By combining eqs. (\ref{bmu2}), we find
\be
B \ = (A_0 - m_{3/2}) \frac{\mu_0}{\mu} + \ \frac{2 m_{3/2}^2}{\mu} K_{12} \ , \label{mubmu}
\ee
that will be used in the next section for phenomenology.
If $K_{12} = 0$, we get $\mu = \mu_0$ and  $B = A_0 -m_{3/2}$,  which is just the familiar mSUGRA relation $B_0 = A_0 - m_0$.

For a suitable choice of gauge kinetic functions $h_{\alpha \beta} =h(\rho) \delta_{\alpha \beta}$,
one generates universal gaugino masses
\be
m_{1/2} \ = \ \frac{\sqrt{2\sigma_0}}{6} \ \overline{D_\rho W(\rho)} \ \partial_{\rho} \ln {\rm Re}\, h \  ,
\ee
where, according to our decoupling hypothesis, we have assumed that $h$ does not explicitly depend on $S$~\footnote{If we allow a coupling of the form $h_A = h_A^0 (1 + \beta_A e^{-2aT} S)$, 
we would find a suppression $m_{1/2} \propto m_{3/2}^2/M_P$.}.
In contrast to the universal scalar masses which are equal to the gravitino mass, $m_{1/2}$ is proportional to $D_\rho W$  and is suppressed by $m_{3/2}/{m_\sigma}$.

As a result, we obtain models resembling
those mediated by anomalies \cite{anom}, where the dominant contributions
to gaugino masses and $A$-terms arise from loop corrections and give \cite{Choi:2005ge}
\be
m_{1/2}^a = \frac{b_a g_a^2}{16 \pi^2} \frac{F^C}{C_0}
\label{anomino}
\ee
and
\be
A_{ijk} = -\frac{\gamma_i + \gamma_j + \gamma_k}{16\pi^2} \frac{F^C}{C_0} \ .
\label{anoma}
\ee
Here $b_a = 11, 1, -3$ for $a=1,2,3$ are the one-loop beta function coefficients, $\gamma_i$ are the anomalous
dimensions of the matter fields $y_i$ and
\be
\frac{F^C}{C_0} = - \frac{1}{3} e^{K/2}  K^{\alpha \bar \beta} K_{\alpha} {\bar D_{\bar \beta} {\bar W}} + m_{3/2} \ \simeq \ m_{3/2}
\ee
is related to the conformal compensator and equals to very high accuracy $m_{3/2}$ in the models we consider.

Because of the loop suppression factor in Eq. (\ref{anomino}),
we are forced to relatively large ($\mathcal{O}$(10-1000) TeV) gravitino masses
in order to have acceptably large gaugino masses\footnote{A posteriori, we know that
for very small $A_0/m_0$ the requirement for relatively large Higgs masses would lead
us to the same conclusion regarding large scalar masses, which to control the relic density would
also require anomaly mediation for gaugino masses.}  Thus, the sparticle spectrum consists of
relatively light gauginos whose masses are determined from anomaly mediation
and large soft scalar masses fixed by the gravitino mass
yielding a spectrum reminiscent of split supersymmetry \cite{split}.
The problem of tachyonic scalars normally associated with anomaly mediated
models is absent here.

In what follows, we will examine the phenomenological consequences of the above model.
In particular, we will see that it is difficult to construct consistent models
if one wants to maintain the possibility of radiative electroweak symmetry breaking.
If the input supersymmetry breaking scale is chosen to be the GUT scale (i.e.
the scale at which gauge coupling unification occurs), one can not choose arbitrarily large
universal scalar masses and insist on a well defined electroweak symmetry breaking vacuum
(i.e., $\mu^2 > 0$).  This difficulty can be alleviated in at least two ways: \\
- increasing the supersymmetry breaking scale, $M_{in} > M_{GUT}$. This is the case that we study in detail in the next section. \\
- Allow for direct couplings between the uplift field(s) $S$ and the Higgs in the Kahler potential, by terms of the type
$S^{\dagger} S H_i^{\dagger} H_i$. In this case, Higgs soft scalar
masses acquire additional corrections proportional to $|F_S|^2$, where $F_S = e^{K/2} K^{S \bar{S} }D_S W$. They are no longer equal to the other scalar masses and are
not necessarily degenerate anymore.  These boundary conditions for scalar masses then resemble
those assumed in non-universal Higgs mass models. This problem can be traced directly back
to our assumption of strong moduli stabilization and small $A$-terms. With large $A$-terms,
there is no difficulty in obtaining electroweak vacuum solutions with large $m_0$ and $\mu^2 > 0$.

Another challenge presented in these types of models
stems from the  mSUGRA relation
between $B_0$ and $A_0$. Unlike CMSSM models \cite{cmssm,eoss}, this relation forces one to
solve for $\tan \beta$ for a given choice of $m_{1/2}, m_0$, and $A_0$.
In the present context, we expect no solutions as there is in effect only a single free
parameter, namely $m_{3/2}$.
However, an interesting extension of minimal supergravity
is one where terms proportional $H_1 H_2$ are added to the K\"ahler
potential as in the Giudice-Masiero mechanism~\cite{gm}.
By introducing a non-minimal coupling to the \K\ potential, one can effectively fix
$\tan \beta$. If in addition, we take $M_{in} > M_{GUT}$, we can in fact formulate
a consistent phenomenological model.

In the next section we briefly review the GM extension to mSUGRA
and the consequences of taking $M_{in} > M_{GUT}$. As a result, we
are forced to consider a specific GUT and here for simplicity, we take minimal SU(5) as
a concrete example.  In section 5, we present the main results of the paper
which include the low energy spectrum as a function of the gravitino mass.
In particular, this amounts to the gaugino and Higgs masses as all of the other supersymmetric
scalars are very heavy. Other phenomenological aspects of the models such as gluino production
at the LHC and the direct and indirect detection of dark matter are discussed in section 6.

\section{GM Supergravity and Super-GUT phenomenology}

As described above, the KL phenomenological model has one free parameter, $m_{3/2}$,
which when extended to include a Giudice-Masiero term, has two free parameters, which we take to be $m_{3/2}$ and $\tan \beta$. This is to be compared with mSUGRA models which have 3 free
parameters or CMSSM models with 4 free parameters.
In the present context, the gaugino masses, scalar masses, and $A$-terms are all determined by the gravitino mass. The alternative of coupling the uplift fields $S$ directly to the Higgs sector, in order to obtain non-universal Higgs masses which are different from the gravitino mass, mentioned in the
previous section, will not be pursued here for simplicity. On the other hand, direct couplings of uplift fields to squarks and sleptons
$ \lambda_{ij} S^{\dagger} S \phi^i \phi_j^{\dagger}$ in the Kahler potential would generically lead to flavor dependence and therefore to FCNC effects. Even for $30-50$ TeV scalar masses, which will be our typical values in what follows, FCNC effects require some degree of degeneracy. This is actually the main phenomenological reason we are imposing no direct couplings between uplift fields and matter fields in our paper. 

From Eq. (\ref{m0}), we expect scalar mass universality at some renormalization scale, $M_{in}$.
In the CMSSM, this scale is usually associated with the GUT scale\footnote{The GUT scale, $\mgut$, is defined as the scale where SU(2) and U(1) gauge couplings
unify and is approximately $1.5\times 10^{16}$~GeV.}. If so, these masses are run down to low
energy using standard renormalization group evolution.
In contrast to the CMSSM, the gravity mediated part of
gaugino masses and $A$-terms in the KL model
are extremely small and their dominant contributions  are determined by
anomalies at any scale using Eqs. (\ref{anomino}) and (\ref{anoma}).
In the CMSSM, $\mu$ and $B$ are solved for
in terms of $m_Z$ and $\tan \beta$:
\ba
\mu^2 & = & \frac{m_1^2 - m_2^2 \tan^2 \beta + \frac{1}{2} \mz^2 (1 -
\tan^2 \beta) + \Delta_\mu^{(1)}}{\tan^2 \beta - 1 + \Delta_\mu^{(2)}} \, ,
\nonumber \\
B \mu  & = & -{1 \over 2} (m_1^2  + m_2^2 + 2 \mu^2) \sin 2 \beta + \Delta_B \, ,
\label{onelooprel}
\ea
where $\Delta_B$ and $\Delta_\mu^{(1,2)}$ are loop corrections~\cite{Barger:1993gh,deBoer:1994he,Carena:2001fw},
and  $m_{1,2}$ are the Higgs soft masses (here evaluated at the weak scale).
In mSUGRA models, however, $B$ can not be determined independently as it must
respect its boundary condition $B_0 = A_0 - m_0$ at $M_{in}$.
Instead, one must solve for $\tan \beta$ (and $\mu$) using the electroweak symmetry breaking conditions \cite{vcmssm,dmmo}. In this sense, the KL models we are describing are more reminiscent of mSUGRA than
the CMSSM.

There are, however, two immediate potential problems with the framework as described:
1) There is no guarantee that reasonable solutions for $\tan \beta$ exist while requiring
$B_0 = A_0 - m_0$ at $M_{in}$. Indeed it is known\cite{vcmssm,dmmo} that only a limited
portion of parameter space (defined by $m_0, m_{1/2}$ and $A_0$) possesses solutions for $\tan \beta$.
2) There is no guarantee that solutions with $\mu^2 > 0$ exist when $m_0$ is very large.
This of course is a well known issue present in the CMSSM. For fixed $m_{1/2}$ and $A_0$,
there is an upper limit to $m_0$ for which there are solutions to (\ref{onelooprel}) with $\mu^2 > 0$
known as the focus point or hyperbolic branch \cite{focus}. This upper limit is also present in
mSUGRA models as well, particularly when $A_0/m_0$ is small (as it is the case under
consideration). As we now describe, neither problem is critical and there are known
and relatively simple solutions to both.

To tackle the problem of $\tan \beta$, consider a Giudice-Masiero (GM) -like contribution to $K$
of the form\cite{gm},
\beq
\Delta K = c_H H_1 H_2  + h.c. \, ,
\label{gmk}
\eeq
where $c_H$ (equal to $K_{12}$ in the previous section)
is a constant, and $H_{1,2}$ are the usual MSSM
Higgs doublets.
The presence of $\Delta K$ affects the boundary conditions for
both $\mu$ and the $B$ term at the supersymmetry breaking input scale, $M_{in}$.
The $\mu$ term is shifted as seen in Eq.~(\ref{bmu2}) to
\beq
\mu_0 + c_H m_0 \, .
\eeq
However,
since we solve for $\mu$ at the weak scale, its UV value is fixed by the low energy
boundary condition. In contrast,
the boundary condition on $\mu B$ shifts from $\mu_0 B_0$
to
\beq
\mu_0 B_0 + 2 c_H m_0^2 \, .
\eeq
We can add the GM term to better connect the solution of the minimization conditions
to a supergravity boundary condition at $M_{in}$. Indeed, by allowing $c_H \ne 0$, we can
fix $\tan \beta$ and derive $\mu$ and $B \mu$ at the weak scale.
By running our derived values of $B(M_W)$ and $\mu({M_W})$ up to the GUT scale,
we can write
\beq
B\mu(\mgut) = (A_0 - m_0)\mu_0(\mgut) + 2 c_H m_0^2 \, ,
\label{gmb1}
\eeq
which is precisely eq. (\ref{bmu2}) of the previous section. Strictly speaking, (\ref{gmb1}) is valid at tree-level in SUGRA and does
not include anomaly contributions. However, the latter are small compared to tree-level values of $m_0$, $B$ and $\mu$, so  (\ref{gmb1}) is an excellent approximation.
In what follows, we  use Eq.~(\ref{gmb1}) to derive the necessary value of $c_H$.

Of course, one must still check, whether the solution for $c_H$ is reasonable (i.e.,
perturbative). In \cite{dmmo}, it was indeed shown that over much of the
mSUGRA parameter space $c_H \lesssim O(1)$. For fixed $\tan \beta$ and $A_0/m_0$,
$c_H$ is reasonably small for most choices of $m_{1/2}$ and $m_0$. Exceptions lying in the
region where $m_{1/2} \gg m_0$ and the lightest supersymmetric particle (LSP)
is the gravitino. When $A_0$ is large, these offending regions are further compressed to small
$m_0$. Thus by allowing non-zero $c_H$, we can always satisfy the mSUGRA boundary
condition for $B_0$ and check a posteriori that $c_H$ is small.

As noted above, in the CMSSM and mSUGRA, there is generally an upper limit to $m_0$
for fixed $m_{1/2}, A_0$, and $\tan \beta$ determined by $\mu^2 = 0$ in Eq. (\ref{onelooprel}).
While it is common to assume that the input supersymmetry breaking scale is equal to the GUT scale,
it is quite plausible that $M_{in}$ may be either below~\cite{eosk} (as in models with mirage
mediation \cite{Choi:2005ge,mixed,llmnr})
or above~\cite{emo,Calibbi,pp,emo3,Baer:2000gf,dmmo} the GUT scale.
Increasing $M_{in}$ increases the renormalization of the soft masses which tends in turn to increase the splittings between the physical sparticle masses~\cite{pp}.  As a consequence,  the focus-point solution for $\mu^2 = 0$ often moves out
to very large values of $m_0$. This feature of super-GUT models is essential for
KL model described here. Note that while the introduction of $M_{in}$ adds a free parameter
to the model, as we will see, our results are very insensitive to the choice of $M_{in}$.
For consistency with the KL paradigm, we should also only consider values of $M_{in} < m_\sigma$.

To realize $M_{in} > \mgut$, we need to work in the context of a specific GUT.
Here, we use the particle content and the
renormalization-group equations (RGEs)
of minimal SU(5)~\cite{pp,others}, primarily for simplicity:
for a recent review of this sample model and its
compatibility with experiment, see~\cite{Senjanovic:2009kr}.
As this specific super-GUT extension of the CMSSM was studied extensively in Refs.~\cite{emo,emo2},
we refer the reader there for details of the model.

The model is defined by the superpotential
\ba
W_5 &=& \muS \Tr\hat{\Sigma}^2 + \frac{1}{6}\lambda'\Tr\hat{\Sigma}^3
 + \mu_H \hat{\calh}_{1} \hat{\calh}_2
 + \lambda \hat{\calh}_{1}\hat{\Sigma} \hat{\calh}_2 \nl
 +({\bf h_{10}})_{ij}
   \hat{\psi}_i \hat{\psi}_j \hat{\calh}_2
 +({\bf h_{\overline{5}}})_{ij} \hat{\psi}_i \hat{\phi}_{j} \hat{\calh}_{1} \, ,
\label{W5}
\ea
where $\hat{\phi}_i$ ($\hat{\psi}_i$) correspond to the $\bf{\overline{5}}$ ($\bf{10}$) representations
of superfields, $\hat{\Sigma}(\bf{24})$, $\hat{\calh}_1(\bf{\overline{5}})$ and
$\hat{\calh}_2(\bf{5})$ represent the Higgs adjoint and five-plets.
Here $i,j=1..3$ are generation indices and we suppress the SU(5) index structure for brevity.
There are now
two $\mu$-parameters, $\mu_H$ and $\muS$,  as well as two new couplings, $\lambda$
and $\lambda^\prime$.  Results are mainly sensitive to $\lambda$ and the ratio of the
two couplings. In what follows, we will fix $\lambda^\prime = 0.1$.

To generalize the GM solution for the $B_0$ boundary condition,
we write
\beq
\Delta K= c_H {\calh_1} {\calh_2} + \frac{1}{2} c_{\Sigma} \Tr \Sigma^2 + h.c. \, ,
\label{gmk2}
\eeq
where ${\calh_{1,2}}$  are scalar components of the Higgs five-plets and $\Sigma$ is the scalar component of the adjoint Higgs.
Thus in principle, we have two extra parameters which can be adjusted to relate
the CMSSM and supergravity boundary conditions for $M_{in}>\mgut$.
Nevertheless, these parameters have virtually no effect on the sparticle mass spectrum
other than allowing us to fix $\tan \beta$ in a consistent manner.

 For $M_{in} > M_{GUT}$, scalar mass universality is defined in terms of the scalar
 components of the Higgses and matter fields in the $\bf{\overline{5}}$ and $\bf{10}$
 representations. At the GUT scale, these must be matched to their Standard Model counterparts.
 More importantly is the matching of the $\mu$ and $B$-terms from SU(5) to Standard
 Model parameters.  These have been discussed extensively in Ref. \cite{Borzumati,dmmo}
 and we do not repeat that analysis here.

There is one aspect of the matching of soft terms at $M_{GUT}$ that is specific to the present model. Dominant contributions to
gaugino masses and $A$-terms are provided by the conformal anomaly
(\ref{anomino},\ref{anoma}), with beta functions and anomalous
dimensions computed with the spectrum at the given energy scale  $E$. For the MSSM  for example, gaugino masses at scale $E$ are given by
\be
m_{1/2}^a (E)= \frac{b_a g_a^2 (E)}{16 \pi^2} \frac{F^C}{C_0}
\ . \label{anomino2}
\ee
Above $M_{GUT}$, on the other hand, we have a unified ($SU(5)$ in our case) theory, with a unified gauge coupling $g_{GUT}$ and a unified beta function $b_{GUT}$. The unified gaugino mass is then given by
\be
{m_{1/2}^{GUT}} (E) \ = \ \frac{b_{GUT} g_{GUT}^2 (E)}{16 \pi^2} \frac{F^C}{C_0} \  \label{anominogut}
\ee
and its value has to be taken into account for the running of soft terms between $M_{in}$ and $M_{GUT}$. However, there is no matching
at $M_{GUT}$ between (\ref{anomino2}) and (\ref{anominogut}). The mismatch is to be interpreted as a threshold effect, due to the decoupling of heavy GUT particles at $M_{GUT}$. The argument is completely similar for the A-terms.

The additional running between $M_{in}$ and $M_{GUT}$ in CMSSM-like models
is very efficient at raising the upper limit on $m_0$ \cite{Calibbi,emo,emo3,dmmo}
provided the Higgs coupling $\lambda$ is sufficiently large.
In mSUGRA-like models, however,
we are still faced with the difficulty of  satisfying the $B_0$ boundary condition and the
GM parameters must be added.
As shown in \cite{dmmo}, for $M_{in} \gtrsim 10^{17}$ GeV, and fixed $\tan \beta$,
$A_0$, $m_0$, and $m_{1/2}$, values of $c_H$ are only small when $\lambda$ is small\footnote{Note that the upper limit on $m_0$ can be raised significantly for small $\lambda$
and large $A_0/m_0$. Of course in the KL models discussed here, $A_0/m_0$ is very small.}.
At large $\lambda$ (needed to raise the upper limit on $m_0$), values of $c_H$ are of order 10
or larger.  However, as also shown in \cite{dmmo}, it is often possible to regulate
$c_H$, by choosing $c_\Sigma \ne 0$ (yet still reasonably small).
Thus, it is possible in the context of a superGUT version of the KL framework,
to obtain a consistent sparticle spectrum.

Let us now summarize the ingredients of the phenomenological model we are considering.
We begin with a no-scale type \K\ potential for the moduli with a racetrack superpotential
as in the KL model. Uplifting is accomplished with an extra heavy Polonyi-like field.
Unlike the KKLT model, here the moduli are superheavy, while
the gravitino remains relatively light.  Scalar mass universality is input at
a renormalization scale $M_{in}$, with $m_0 = m_{3/2}$ at that scale.
Gravity mediation supplies vanishingly small gaugino masses of order $m_{3/2}^2/m_\sigma$
and $A$-terms of order $\Lambda^2 m_{3/2} \ll m_{3/2}$,
and thus these quantities receive their
dominant contribution from anomalies. As a result, we have $m_{1/2}/m_0 \ll 1$
as well as $A_0/m_0 \ll 1$, and choose $M_{in} > M_{GUT}$ to allow
solutions to the EWSB equations with $\mu^2 > 0$. In addition,
we introduce GM parameters, $c_H$, and $c_\Sigma$ so that $\tan \beta$
can be held fixed and still satisfy the boundary condition for $B_0$.
We will see that for $c_\Sigma \gtrsim -1$, we can obtain $c_H$ close to 0.
We expect that the alternative road we mentioned, of imposing boundary conditions at $M_{\rm in} = M_{\rm GUT}$, and changing
Higgs scalar masses via direct couplings of uplift fields to the Higgs sector, will lead to a similar phenomenology.

\section{Low-energy spectra}

We next describe the resulting sparticle spectrum and the predicted low energy phenomenology we
expect from the KL motivated class of models.
As should be clear from the preceding discussion, below the superheavy scale
(GUT fields and moduli), there are sparticles with masses of order $m_{3/2}$,
and those with much lighter masses as determined from anomalies.

Because of the boundary condition $m_0 = m_{3/2}$, all of the scalar partners of
the matter fields have mass close to $m_{3/2}$.  For example, let us take
$m_{3/2} = 32$ TeV, with $\tan \beta = 25$, $M_{in} = 5 \times 10^{17}$ GeV, and $\lambda  = 1.35$.
By choosing $c_\Sigma \simeq -0.85$, we can obtain $c_H = 0$ for this point.
Not surprisingly, all of the first and second generation sfermions have masses close to $m_{3/2}$.
The lighter stau lepton has a mass of 29.6 TeV, and the lighter stop and sbottom quarks have
masses of 24.2 and 26.9 TeV.  All of which are past the current (and future) reach of the LHC.
The Higgs soft masses run considerably and both mass-squared are negative with
$|m^2|^{1/2} = $ 10.9 and 25.1~TeV. The $\mu$ parameter from Eq. (\ref{onelooprel}) is also
very large (20.4 TeV) unlike the assumption made in many models of split supersymmetry.
As a result, the pair of Higgsinos and the Higgs-like chargino have large masses (22.0 TeV).
Similarly, the heavy Higgs scalar and pseudo-scalar have a mass of 20~TeV.
These results are not very sensitive to any of the assumed input parameters and to a
first approximation are all close to $m_{3/2}$. The full spectrum for this point and other
test points are given in Table \ref{tab:bm}.

In contrast, the gauginos and one (wino-like) chargino remain relatively light.
Their masses are dominated by the anomaly contributions given in Eq. \ref{anomino}.
For the specific choice of parameters adopted above, we have a nearly degenerate
neutral and charged pair of winos at 107 GeV (the neutral wino is the LSP) and
a 314 GeV bino.  The gluino mass is 1.0 TeV. Note that as we use two-loop
expressions for the running of gaugino masses, our results differ slightly
from the approximate values given in Eq. (\ref{anomino2}).

As we saw, the supersymmetry breaking scalar masses at the weak scale are
$M_{SUSY}\simeq m_{3/2}\sim 10-1000$~TeV, while the gauginos are
significantly lighter. Such heavy scalars lead to important quantum
corrections in the Higgs sector. These are enhanced by large logarithms of the type
$\log (M_{weak}/M_{SUSY})$. In fact, most of the public codes assume
$M_{SUSY}$ less than a few TeV, otherwise the computations become
unreliable. Thus in the computation of the light CP-even Higgs boson mass,
$m_h$, we follow the procedure described in Refs.\cite{mhsplit,gs} and 
we add in the contributions due to sbottoms-induced corrections, though these
are small in the models considered.
The Higgs mass is effectively given by its quartic coupling
\be
m_h^2 = 2 v^2 \lambda_H(Q) \, ,
\ee
where $v$ is the Higgs vacuum expectation value and we evaluate the quartic coupling at $Q=m_t$.
The tree-level quartic coupling is simply
\be
\lambda_H = \frac{1}{4}(\frac{3}{5} g_1^2 + g_2^2) \cos^2 2 \beta
\ee
at the scale where heavier higgs fields decouple.
The masses of
the heavy CP-even and CP-odd Higgs bosons are of order $M_{SUSY}$ and thus can be reliably
computed by conventional techniques.

The heavy scalars decouple at approximately the scale, which we take to be
$M_{SUSY}=\sqrt{m_{\stop_1}m_{\stop_2}}$, and the low-energy theory
contains only SM fermions, gauginos,  and one SM-like
higgs doublet. Thus, in the effective theory below $M_{SUSY}$,
in addition to the regular SM gauge, Yukawa and Higgs
quartic couplings, we have Yukawa-like Higgs-higgsino-gaugino couplings.
Since SUSY is broken below $M_{SUSY}$ those
Yukawa-like couplings are no longer equal to corresponding gauge couplings
and also renormalize differently. We perform RGE evolution of all
effective-theory couplings at the 2-loop level and take into account
1-loop threshold effects, thus obtaining the Higgs mass, $m_h$ at full
next-to-leading order accuracy.

\begin{figure}
\begin{center}
\epsfig{file=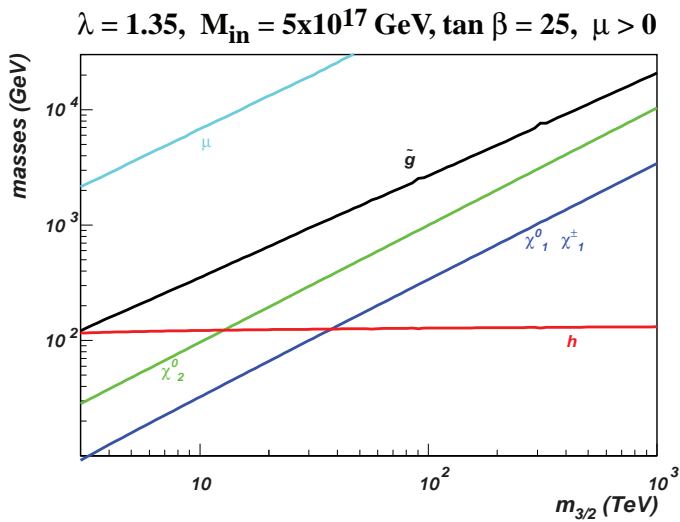,height=7cm}
\end{center}
\caption{\it
The gaugino and chargino masses and the $\mu$-term as a function of the gravitino mass, $m_{3/2}$.
Here we have chosen, $\tan \beta = 25$, $M_{in} = 5 \times 10^{17}$ GeV, $\lambda = 1.35$.}
\label{fig:masses}
\end{figure}

The sparticle spectra for a few values of $m_{3/2}$ are given in Table \ref{tab:bm}.
The (light) fields have masses which scale with the gravitino mass.
In Fig.~\ref{fig:masses}, we show the resulting masses of the three gauginos, chargino
(degenerate with the wino-like gaugino) and Higgs mass as a function of $m_{3/2}$.
For reference, we also plot $\mu$.
For this example, we have chosen the same input parameters discussed above, namely
$\tan \beta = 25$, $M_{in} = 5\times 10^{17}$~GeV, and $\lambda = 1.35$.
In addition, here and in all subsequent figures, we have taken
$\lambda^\prime = 0.1$ and $c_\Sigma = -0.85$.
For the range of $m_{3/2}$ shown, $c_H$ can be made to vanish for
$c_\Sigma = -0.85$ - $-0.91$.
We emphasize that these parameters do not affect the mass spectrum, but
allow us to take fixed $\tan \beta$. As one can see the ratio between the gaugino masses
is relatively fixed as one might expect from the anomaly conditions. Any deviations from
this are a result of two loop effects. Notice that on this vertical scale, the Higgs mass is nearly constant.

\begin{table*}
\begin{center}
\begin{tabular}{lcccccc}
\hline
parameter & 1 & 2 & 3 & 4 & 5 \\
\hline
$m_{3/2}$ [TeV]  & 32 & 50 & 100 & 500 & 1000 \\
\hline
$\mgl$ [TeV]      & 1.0 & 1.5 & 2.7 & 11.1 & 20.8 \\
$\msf\chi 1$ [GeV]& 107 & 168 & 338 & 1705 & 3423\\
$\msf\chi 2$ [GeV]& 314 & 495 & 1000 & 5130 & 10400\\
$\msf\chi 3$ [TeV]& 22.0 & 34.9 & 70.7 & 367 & 745\\
$\msf\chi 4$ [TeV]& 22.0 & 34.9 & 70.7 & 367 & 745\\
$\mchr 1$   [GeV]& 107 & 168 & 338 & 1705 & 3420 \\
$\mchr 2$   [TeV]& 22.0 & 34.9 & 70.7 & 367 & 745 \\
$\msf t 1$ [TeV]  & 24.2 & 38.0 & 77.2 & 397 & 803\\
$\msf t 2$ [TeV]  & 26.8 & 42.1 & 84.6 & 428 & 860\\
$\msf b 1$ [TeV]  & 26.9 & 42.1 & 84.7 & 428 & 860\\
$\msf b 2$ [TeV]  & 30.6 & 47.9 & 96.0 & 483 & 969\\
$\msf q L$ [TeV]  & 31.4 & 49.2 & 98.5 & 494 & 990\\
$\msf u R$ [TeV]  & 31.5 & 49.3 & 98.7 & 495 & 990\\
$\msf d R$ [TeV]  & 31.6 & 49.4 & 98.9 & 496 & 992\\
$\msf\tau 1$ [TeV] & 29.6 & 46.2 & 92.3 & 459 & 917\\
$\msf\tau 2$ [TeV] & 31.2 & 48.7 & 97.5 & 488 & 978\\
$\msf\nu \tau$ [TeV]& 31.2 & 48.7 & 97.5 & 488 & 978\\
$\msf e L$ [TeV]  & 31.9 & 49.8 & 99.6 & 498 & 996 \\
$\msf e R$ [TeV]  & 32.0 & 50.0 & 100 & 500 & 1000 \\
$\msf \nu L$ [TeV]  & 31.9 & 49.8 & 99.6 & 498 & 996 \\
$m_h$ [GeV]       & 125 & 127 & 128 & 132 & 133 \\
$\mu$ [TeV]       & 20.4 & 32.3 & 65.0 & 333 & 673 \\
$m_A$ [TeV]       & 19.5 & 30.6 & 58.4 & 262 & 494 \\
\hline
$\Omega_{\schi}h^2$ & 0.0003 & 0.0008 & 0.0030 & 0.067 & 0.26 \\
$\sigma^{SI}(\chi_1 p)\times 10^{14}$ [pb] & 4.74 & 1.81 & 0.44 & 0.02 & 0.003 \\
$\sigma^{SD}(\chi_1 p)\times 10^{12}$ [pb] & 6.78 & 0.94 & 0.04 & 0.0008 & 0.001\\

\hline
\end{tabular}
\caption{Input parameters and resulting masses and rates
for benchmark points with $M_{in}=5\times 10^{17}$~GeV, $\lambda=1.35$, $\lambda'=0.1$, $c_\Sigma = -0.85$, $\tan\beta=25$, $\mu > 0$ and $m_t=173.1$~GeV.
}
\label{tab:bm}
\end{center}
\end{table*}

Note that the LEP bound on the chargino mass of 104 GeV, provides a direct lower limit to
$m_{3/2}$ in these models. As can be seen from Fig.~\ref{fig:masses}, this chargino mass
limit implies $m_{3/2} \gtrsim 31$ TeV. The Higgs mass at this value of $m_{3/2}$ is 125.3 GeV
(at $\tan \beta = 25$).
Thus, the limit on the chargino mass directly implies a lower limit on the Higgs mass of roughly
125 GeV in amazing agreement with the recent discovery claimed at the LHC \cite{LHCHiggs}.
Furthermore, at the lower bound on $m_{3/2}$, the gluino mass is 970 GeV
close to the current lower bound from the LHC \cite{lhcgluino}.

Because, the gaugino and chargino spectrum is determined by anomalies, it
is almost completely independent of the choice of the parameters chosen in Fig \ref{fig:masses}.
We have explicitly verified that for several choices of $\tan \beta$  and $\lambda$,
and a range of input scales,
$M_{in}$, up to the Planck scale, differences in the mass spectra are negligible.

\begin{figure}
\begin{center}
\epsfig{file=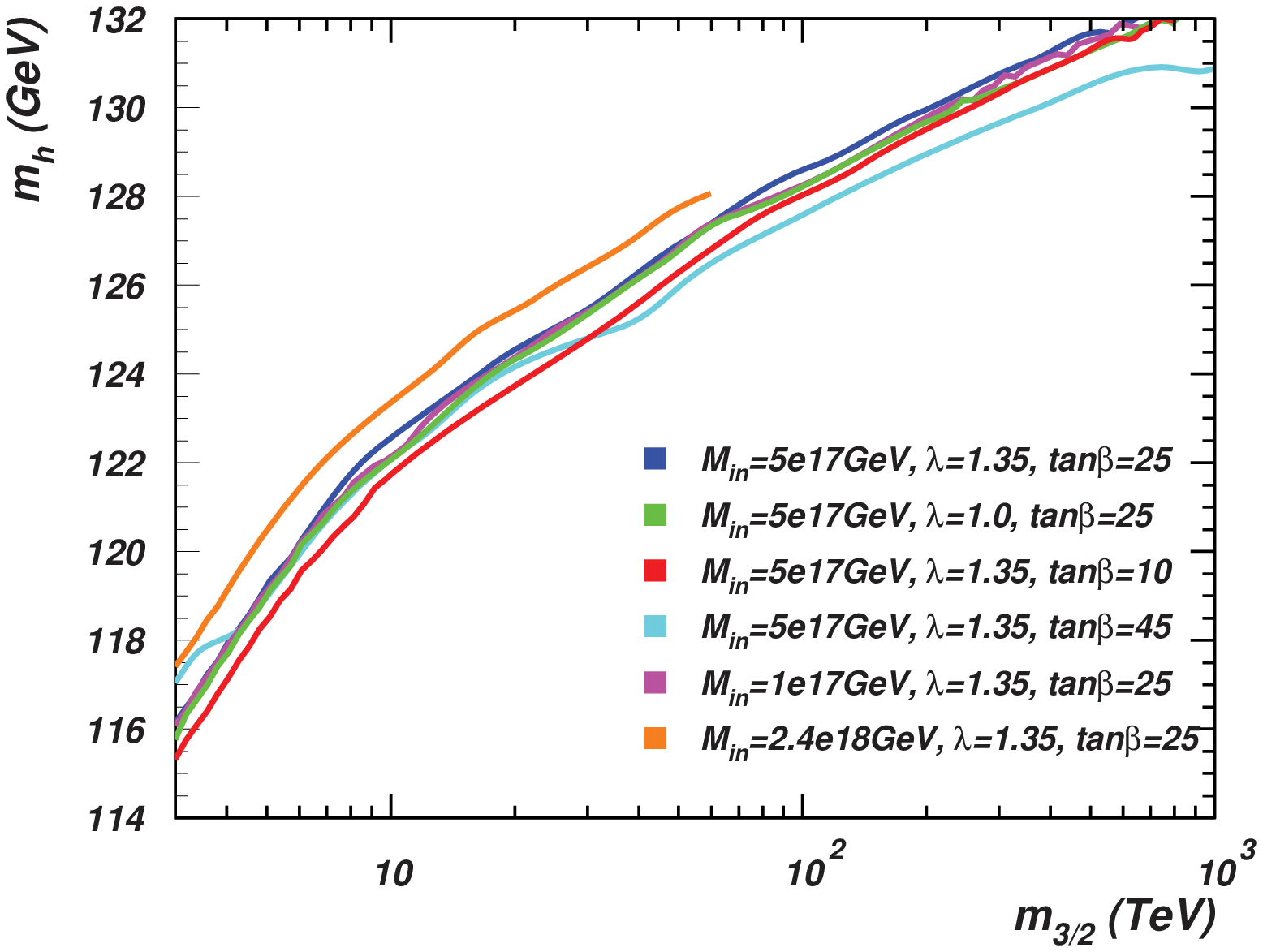,height=7.0cm}
\end{center}
\caption{\it
The Higgs mass as a function of the gravitino mass, $m_{3/2}$.
Here we have chosen, several combinations of $\tan \beta$, $M_{in} $, and $\lambda$
as indicated on the figure.}
\label{fig:Higgs1}
\end{figure}

\begin{figure}
\begin{center}
\epsfig{file=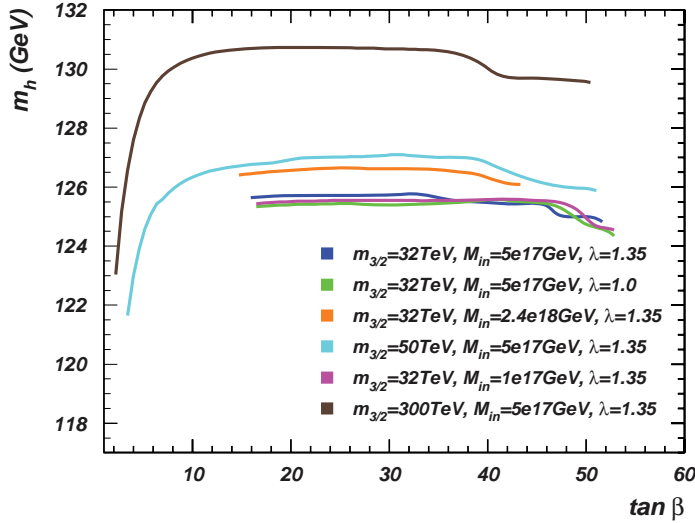,height=7.0cm}
\end{center}
\caption{\it
The Higgs mass as a function of $\tan \beta$.
We have chosen, several combinations of $m_{3/2}$, $M_{in} $, and $\lambda$
as indicated on the figure.}
\label{fig:Higgs2}
\end{figure}

The exception is the Higgs mass which we show separately in Figs.~\ref{fig:Higgs1} and \ref{fig:Higgs2}.
In Fig.~\ref{fig:Higgs1}, we show the Higgs mass as a function of $m_{3/2}$ for several sets of input parameters. In addition to our base model with $\tan \beta = 25$, $M_{in} = 5 \times 10^{17}$ GeV,
$\lambda = 1.35$, we vary each of these to show the behavior of the Higgs mass as a function of $m_{3/2}$.
For the most part, we see that they are largely insensitive to the choice of input parameters
and all yield masses between 125~GeV (when $m_{3/2} > 30$ TeV so that $m_{\chi^+} > 104$~GeV)
and $\sim 130$ GeV~\footnote{We note that there is at least a 2-3 GeV uncertainty in the Higgs mass calculation,
so that we can not out of hand exclude results with Higgs masses at 130 GeV.}
Indeed the Higgs mass determination allows one to set an upper limit on the split susy scale \cite{gs,split limit}
or $m_{3/2}$ in the present context.

In Fig.~\ref{fig:Higgs2}, we show the Higgs mass as a function of $\tan \beta$. For $\tan \beta \gtrsim 10$,
the Higgs mass is relatively insensitive to $\tan \beta$ and decreases for smaller values of $\tan \beta$.
Several of the curves are cut off at low $\tan \beta$ when the chargino mass falls below the LEP limit.
We see again that the results are also insensitive to the particular choice of model for
$m_{3/2}$ = 32~TeV, but increases with $m_{3/2}$ as we saw in Fig.~\ref{fig:Higgs1}.

Note that the curves in Fig.~\ref{fig:Higgs2} terminate at large $\tan \beta$ when solutions to the 
electroweak symmetry breaking conditions can no longer be obtained. 
This is due to the effect of the bottom (and to a lesser degree, tau) Yukawa coupling on the RGE 
evolution of Higgs mass
parameters. At these values of $\tan \beta$, the bottom Yukawa coupling is large and comparable in magnitude to the top Yukawa coupling. Since Yukawa terms tend to decrease scalar masses, $m^2_{1}$ at the weak scale is driven to smaller and even to negative
values. At some point it becomes impossible to find solutions to Eqs. (\ref{onelooprel}).
A similar decrease in $m^2_{1}$ is found when both $M_{in}$ and $m_{3/2}$ become sufficiently large~\cite{emo}, as can be seen from the
behaviour of the orange curve in Fig.~\ref{fig:Higgs1} that terminates at $m_{3/2}\simeq 60$~TeV.


\section{Other phenomenological aspects}

\subsection{Gluinos}

Multi-TeV scalars have severe consequences for the low energy spectra and
LHC discovery prospects.
It seems obvious that except for a hypothetical SLHC extension, the scalar
sector of constructions
such as the one described above
is unreachable at the LHC. Nevertheless, the gaugino spectrum is
relatively light
(reduced by a loop factor
generated by anomaly mediation) and a 1 TeV gluino can still be detectable.
However,  the classical 2-body decay mode $\tilde g \rightarrow \tilde q
q$ is kinematically forbidden.
The dominant decay modes become:

\noindent
1)  The three-body decay $\tilde g \rightarrow q \tilde q^* \rightarrow q
q \schi^0_1$ (through the exchange of
a virtual squark $\tilde q^*$) with a 2 jet plus missing $E_T$ signature.

\noindent
2) The 2-body decay modes $\tilde g \rightarrow g \schi^0_1$ generated by
squark-quark loop diagrams.

\noindent
There have been several analyses of gluino decay in split supersymmetry
\cite{gldecay,Sato:2012xf}.
These showed that as long as the scalar masses are below $\sim 10^4$ TeV,
the lifetime of the gluino
is too short to be detected.
Since the two body decays occur primarily via the Higgsino component of the
neutralino \cite{btw}, these will be highly suppressed in the models under consideration
with $\mu \gg m_{1/2}^a$ and a negligible Higgsino component in the either of the two
light neutralinos. While there is a gaugino component to the
2-body decay, it cancels in the limit of degenerate squark masses
due to the tracelessness of the diagonal generators of the electroweak gauge group.
Since in our scenario, squarks are of order 10-1000TeV and
close in mass, the 2-body loop decay is always negligible.

The three body gluino decay rate to winos can be written as \cite{Sato:2012xf}
\begin{eqnarray}
&&
\Gamma(\tilde g \rightarrow q \bar q \widetilde \chi) \ \approx \
\frac{9}{4} \frac{g^2_3 g_2^2}{768 \pi^3} \frac{m^5_{\tilde g}}{m^4_{\tilde q}} \, .
\end{eqnarray}
Assuming that we are sufficiently far from thresholds (which we are in
the cases under consideration), the above expression is a reasonable approximation.
Numerically, however, the decay rates for $m_{3/2} < 1000$ TeV are sufficiently short
so as to make the displaced vertices in gluino decays unobservable at the LHC.
For the benchmark points above, the decay rates range from $5.3\times 10^{-11}
- 2.1 \times 10^{-10}$ TeV for $m_{3/2} = 32 - 1000$ TeV.
This corresponds to a decay length of 3.7 -- 0.9 nm, respectively.
These lengths are far below the LHC resolution of order 10 $\mu$m.

Nevertheless,  gluinos can be observed at the LHC since their pair production cross section is above the fb level for $m_{\tilde g} \lesssim 1.8$~TeV ~\cite{Baer:2007ya}. Gluinos can be searched for using the usual multi-jet plus missing energy analysis. It was shown that the gluino-wino mass difference can be determined from the endpoint of the di-jet invariant mass distribution with the accuracy of 5\% ~\cite{asai}. Additional information about the gluino mass can be obtained from the measurements of the total gluino production cross section~\cite{Baer:2007ya} or from the effective mass $M_{eff}$ distribution~\cite{meff}.

\subsection{Charginos}
A related possibility for accelerator detection is through the observation of charged tracks of winos, as was suggested for AMSB-type gauginos~\cite{ggw}. When $\mu$ is much larger than $M_2$, the lighter chargino $\schi_1^\pm$ is mostly wino and is only slightly heavier than the wino-like LSP.
The chargino decay rate depends strongly on the $\schi_1^\pm - \schi_1^0$ mass splitting: if the splitting is larger than the mass of the charged pion, the dominant decay is $\schi_1^\pm \rightarrow \schi^0 + \pi^\pm$. The rate of this decay is quite slow and the chargino lifetime is of order $10^{-10}$s. The resultant pion is very soft and one will only see the charged track stub, i.e. an abruptly terminating charged track, from $\schi_1^+$. That track stub is only a few centimeters long for the roughly 200 MeV mass gap we have, so it will be difficult to observe at LHC. However, it has been argued that the track stubs could be observed in a specific mono-jet searches, provided that wino pairs are produced in the association with a hard jet that will serve as a trigger~\cite{ggw,ctchar}.

\subsection{Dark Matter}

\begin{figure}
\begin{center}
\epsfig{file=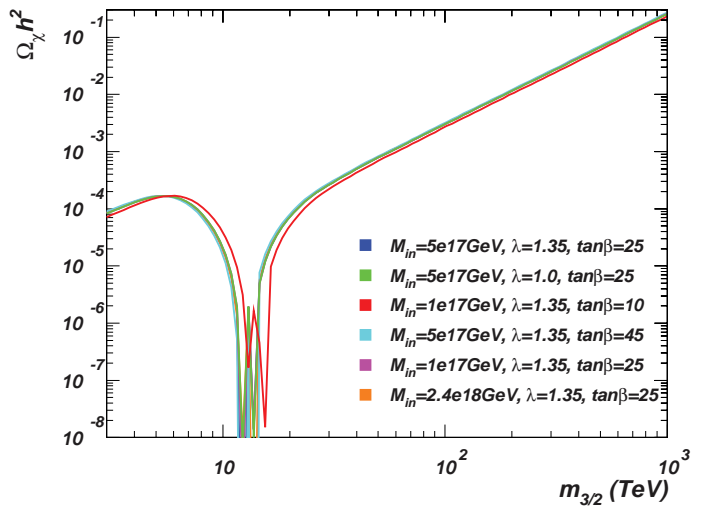,height=6.7cm}
\end{center}
\caption{\it
The LSP (neutral wino) relic density, $\Omega_\chi h^2$, as a function of the gravitino mass, $m_{3/2}$.
Here we have chosen, several combinations of $\tan \beta$, $M_{in} $, and $\lambda$
as indicated on the figure.}
\label{fig:relic}
\end{figure}

Since the only `light' sparticles in the spectrum are the neutral gauginos and charged wino,
and since these sparticle masses are fixed by anomalies, it is easy to see that
the LSP is a wino. As a consequence, the relic density of LSPs is in general relatively small.
The neutral and charged winos are nearly degenerate and the annihilations and co-annihilation
cross sections to $W^+W^-$ are large. Thus for our reference point at $m_{3/2} = 32$ TeV and
$\tan\beta = 25$, the relic density of winos is $\Omega_\chi h^2 = 2.8 \times 10^{-4}$, far below
the density ascertained by the cosmic microwave background anisotropy spectrum \cite{wmap}.
In Fig.~\ref{fig:relic}, we show the wino relic density as a function of $m_{3/2}$.
As one can see, for values of $m_{3/2}$ near the lower bound of 30 TeV,
the relic density is far too small to account for the dark matter density inferred from the CMB.
There are two noticeable dips at $m_{3/2} = 12 - 13$ TeV,  which occur
in all cases considered. These are due to the strong
co-annihilation between the nearly degenerate neutral and charged winos. The first dip occurs
because of the s-channel $W$-exchange in neutral and charged winos when
$m_\chi \simeq m_{\schi^{\pm}} \simeq m_W/2$, whilst the second dip occurs at slightly
higher $m_{3/2}$ because of s-channel $Z$-exchange in chargino annihilation
when $m_\chi \simeq m_{\schi^{\pm}} \simeq m_Z/2$.  Of course, these chargino
masses are well below the LEP bound and the value of $m_{3/2}$ where the dips occur is
excluded.

We must therefore, address the question of dark matter in the context of this class of phenomenological
models.  Three possibilities are relatively straightforward:

1) The dark matter is something else other than a wino (e.g. an axion).
This is a rather obvious possibility. While supersymmetric dark matter is an attractive possibility,
it is not unique and axions present us with another well motivated candidate.
Originally it was assumed that axions give a proper contribution to dark matter only if their mass is $O(10^{-5})$ eV \cite{Abbott:1982af}.
However, later it was found that if the PQ symmetry was broken during inflation, and the Hubble constant during the last stage of inflation is sufficiently small, which is necessary to suppress isocurvature axion perturbations, then anthropic considerations lead to a desirable axionic contribution to dark energy for a broad range of axion masses $m_{a} \ll10^{-5}$ eV \cite{Linde:1987bx}. One may also consider models with many light axion fields, the axiverse \cite{Arvanitaki:2009fg}.

2) LSPs are created non-thermally via gravitino or moduli decay.
In principle, moduli decay to gravitinos which subsequently decay to
neutralinos can be quite dangerous, as their late decays may often overproduce
LSPs leaving one with a relic density in excess of the WMAP determination.
In our case, the modulus of concern is the Polonyi-like field $S$.
However, because of the structure of the \K\ potential in Eq. (\ref{kalpol}),
we expect the field $S$, to be pinned to the origin during inflation, when its
mass is large ($H^2/\Lambda$ \cite{klor,Kallosh:2010ug,Kallosh:2010xz}) and subsequent oscillations to the final minimum of
$S$ will have an amplitude no larger than $\Lambda^2$, thus greatly suppressing the
energy density in its oscillations. For $\Lambda < 10^{-3} - 10^{-4}$, these oscillations
and decays become harmless. Furthermore, if the adiabatic relaxation mechanism in the models with small $\Lambda$
is operative \cite{adrel}, then the amplitude of oscillations is exponentially suppressed, and therefore the contribution of the decay products of the Polonyi field to the LSP relic density is exponentially small. Thus, strong moduli stabilization may play an additional important role in this context.
While the complete evolution of these moduli
in the context of inflation is interesting for further study, here we will assume that because
of the suppression in moduli oscillations, they will not contribute significantly to the relic density
of dark matter. On the other hand, since suppression of the LSP production as a result of the Polonyi field decay is sensitive to the choice of $\Lambda$, one might find such values of this parameter where this effect may lead to a significant increase of the dark matter density.

Instead of moduli decay, we focus on the well known fact that gravitinos are produced during reheating after inflation~\cite{EKN,Kawasaki:1994af,enor,buch}. The calculation of Ref.~\cite{buch} yields
\begin{equation}
 {n_{\tilde G} \over n_\gamma} \; = \; 1.2 \times
10^{-11}
\left( 1 + {m_{\widetilde g}^2 \over 12 m_{3/2}^2}  \right) \times
\left(
{T_R
\over 10^{10}~{\rm GeV}}
\right) \, ,
\label{gravY}
\end{equation}
where $T_R$ is the
reheating temperature  achieved at the end of the
inflationary epoch.

Since $m_{\widetilde g} \ll m_{3/2}$ in the models considered here, we can ignore
the middle term in Eq. (\ref{gravY}), which can be rewritten as
\be
\Omega_{3/2} h^2 \simeq 0.4 (\frac{m_{3/2}}{\rm TeV}) (\frac{T_R}{10^{10} {\rm GeV}}) \, .
\ee
Then the relic density of winos is given simply by
\be
\Omega_\chi h^2 = \frac{m_\chi}{m_{3/2}} \Omega_{3/2} h^2 =
0.4 (\frac{m_{\chi}}{\rm TeV}) (\frac{T_R}{10^{10} {\rm GeV}})  \, .
\ee
Thus for a wino mass of about 100 GeV, and a reheat temperature
of $T_R = 3 \times 10^{10}$ GeV, we obtain the correct dark matter relic density.

Typically, reheating temperatures as large $O(10^{10})$ GeV are excluded from
big bang nucleosynthesis (BBN) \cite{nos,EKN,enor,buch,others2,kmy}.
These limits are obtained from
the effects of gravitino decay on the background of light elements produced during or after BBN.
The gravitino decay rate can be written as
\be
\Gamma_{3/2} = \frac{m_{3/2}^3}{4\pi} \, ,
\ee
for example, for a decay to a gluon and gluino,
corresponding to a lifetime
\be
\tau_{3/2} \simeq 3 \times 10^4 {\rm s} \left( \frac{1 {\rm TeV}}{m_{3/2}} \right)^3  \, .
\ee
Clearly, for gravitino masses of order 100 GeV, the lifetime is of order
$10^8$ s and can affect the light element abundances. However,
for gravitino masses in excess of 30 TeV, the lifetime is of order 1 s
and thus gravitinos decay before BBN and there is practically no limit on the reheat temperature. A problem appears only if gravitinos may decay to LSPs, which may give an additional contribution to dark matter. But this is exactly what we want.

While the reheat temperature may not be constrained by BBN, reheating after inflation
in KL-type models requires some attention. In Ref. \cite{klor}, an examination of possible inflaton decay
channels in KL supergravity showed that reheating is in general suppressed as in
the cases of no-scale supergravity \cite{ekoty} or racetrack  inflation \cite{BlancoPillado:2004ns}.
However a decay channel to gauge bosons is possible if the gauge kinetic term, $h_{\alpha\beta}$ has a non-trivial coupling
to the inflaton $\phi$. Expanding $h_{\alpha \beta}$ in terms of $\phi$, we can write
\be
 h_{\alpha \beta} = \bigl(h(\rho) + d_{\phi} \phi\bigr) \delta_{\alpha \beta} .
 \label{hab}
\ee
As written, $h$ is a function of the moduli (normally included for the generation of gaugino masses - though
these are greatly suppressed in the present context) and the inflaton, $\phi$. The coupling $d_\phi$ is a constant.
This coupling was shown to lead to a reheating temperature of  order
\be
T_R \sim   d_{\phi} \times 10^{9}\  {\rm GeV} \ .
\ee
for $m_\phi = 6 \times 10^{-6}$ in Planck units.
Thus barring other possible enhancements, we would require $d_\phi \gtrsim 10$
to achieve reheating large enough to obtain the correct relic density of neutralinos
after gravitino decay.

A similar situation appears if the inflaton field is a pseudo scalar coupled to vectors as $~ d_{\phi}\phi F\tilde F$. Such models, with $d_{\phi} \sim 10^{2}$ and large reheating temperature, were recently considered in \cite{Barnaby:2010vf} as a potential source of non-Gaussian inflationary perturbations.

3) We increase $m_{3/2}$  sufficiently (to $\gtrsim 500$ TeV) to get the right relic density.
As is clearly seen in Fig.~\ref{fig:relic}, for suitable $m_{3/2} \simeq 650$ TeV,
we can obtain the WMAP density
of $\Omega_\chi h^2 \simeq 0.11$. Of course at this value of $m_{3/2}$, we obtain $m_h \simeq 128.5$ GeV slightly in
excess of the recent LHC measurement.
However, as we noted earlier, there is certainly some uncertainty in the calculated value
of $m_h$ and so we can not entirely exclude this possibility.


\subsection{Direct and indirect detection of dark matter}

\begin{figure}
\begin{center}
\epsfig{file=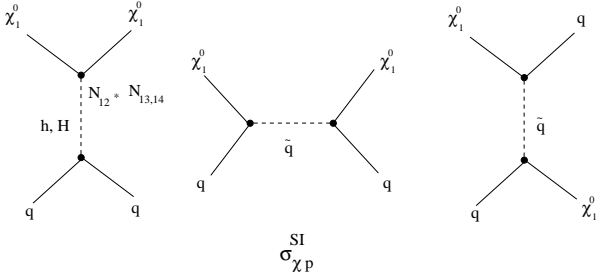,height=3.6cm}
\end{center}
\caption{\it
Direct detection processes for the neutralino-nucleon elastic scattering.}
\label{fig:direct}
\end{figure}

\begin{figure}
\begin{center}
\epsfig{file=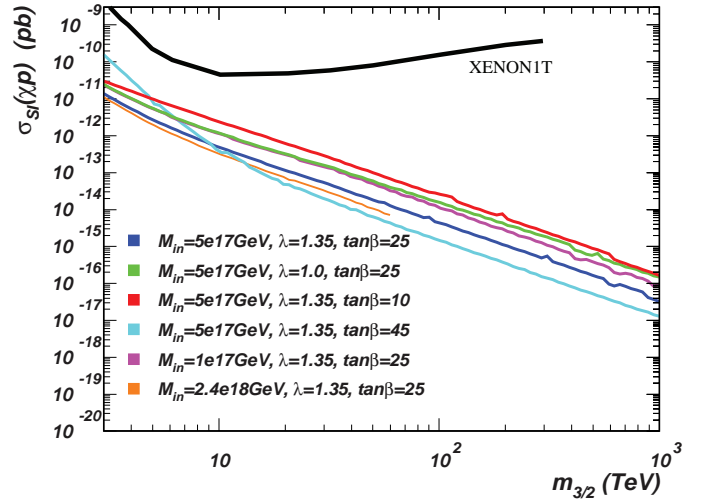,height=6.7cm}
\end{center}
\caption{\it
The spin independent elastic cross section, $\sigma_{\chi p}$ , as a function of the gravitino mass, $m_{3/2}$. Also shown is the projected limit for a XENON-1 ton detector \cite{xenon}.}
\label{fig:sigmasi}
\end{figure}

A wino--like dark matter candidate has several features which makes its direct detection modes, through the measurement
of its scattering off a nucleus, difficult to observe. Indeed the two main scattering modes are the t-channel Higgs exchange (h or H)
and the s- and t-channel squark exchange (see Fig.\ref{fig:direct}). The
t--channel SM Higgs exchange  is strongly suppressed due to the coupling
of the LSP to the Higgs boson. Indeed, this coupling is proportional to the product of the Higgsino and
gaugino components of the neutralino. With such a heavy Higgsino ($\simeq 20-30$ TeV), the lightest neutralino (wino)
has a very small higgsino component (much less than 1 \%), thus reducing the effective $\schi^0_1~\schi^0_1~q \bar{q}$ coupling.
Processes with heavy Higgs exchange or squark exchange (Fig.\ref{fig:direct}),
 are both also strongly suppressed with such a heavy scalar spectrum ($\simeq 30$ TeV),
 giving a reduction
by a factor $10^3$ to $10^4$ compared to typical WIMP interactions on nucleon. This is clearly illustrated in Fig.
 \ref{fig:sigmasi} where a typical cross section for a 100 GeV neutralino ($m_{3/2}\simeq 30$ TeV) is
 $10^{-14}$ pb whereas classical WIMP interaction should lie between $10^{-8}$ to $10^{-12}$ pb.
 This result is relatively independent of $\tan \beta$ or $M_{in}$ as the arguments developed above
 are quite general.
 For reference, the anticipated reach of a XENON 1 ton detector is about a few $\times 10^{-11}$
 pb  \cite{xenon} and is shown by the black curve.

\begin{figure}
\begin{center}
\epsfig{file=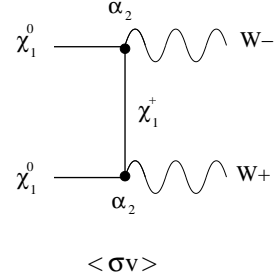,height=3.6cm}
\end{center}
\caption{\it
Main neutralino annihilation channel for indirect detection constraint imposed by FERMI.}
\label{fig:indirect}
\end{figure}

\noindent
On the other hand, it is well known that a neutralino with a dominant wino component has a large
s-wave annihilation cross section, which implies possibilities for indirect signals. The main annihilation channel
is the t-channel exchange  of the chargino (see Fig. \ref{fig:indirect}). In the anomaly  mediation scenario,
the mass degeneracy between $\schi^0_1$ and $\schi^+_1$ together with the relatively strong $SU(2)$ coupling
 generates a high rate of $W^+ W^-$ final states
(around 80 \%). However, such final states are strongly constrained by the recent analysis of dwarf galaxies
by the FERMI telescope \cite{FERMI}. Due to the lower limit on the chargino mass from LEP constraint,
this is also the lower limit for the lightest neutralino as they are nearly degenerate. For the benchmark point
$m_{3/2}=32$ TeV giving $m_{\schi^0_1}=107$ GeV, we obtained
$\langle \sigma v \rangle = 3.5 \times 10^{-7} \mathrm{GeV^{-2}} \simeq 4 \times 10^{-24} \mathrm{cm^{-2} s^{-1}}$
which corresponds to the derived 95\% CL upper limit of FERMI \cite{FERMI} in the case of $W^+ W^-$ final state.
Moreover, as the dependence of the fluxes on the mass is proportional to $1/m_{\chi}^2$, and the limit
of FERMI is even less constraining for heavier DM masses, we can safely conclude that the combined LEP/FERMI data
do not affect the parameter space considered.

\section{Conclusions}
\label{sec:concl}

The Polonyi, or more generally, the moduli problem~\cite{Polonyi} has been a persistent problem
for supergravity models put in a cosmological context. These problems were compounded
in KKLT models of moduli stabilization as they place a severe requirement
on the Hubble parameter during inflation, $H \lesssim m_{3/2}$, in order to
avoid destabilization of the volume modulus. The mass of the volume modulus is relatively small, so even if one manages to avoid the volume modulus destabilization and the resulting decompactification, one may encounter a version of the cosmological moduli problem related to this field. Furthermore, if $F$-term supersymmetry breaking is used to uplift the AdS supersymmetric vacuum in KKLT, one may
have a classic Polonyi problem in addition to the cosmological problems associated
with the string theory moduli. Depending on its mass, the gravitino may also present
a plethora of problems.

On the other hand, a relatively simple modification of the KKLT superpotential
allows for strong moduli stabilization.  In the particular example of a racetrack superpotential,
as in the KL model, the volume modulus may be very heavy ({\it i.e.}, GUT scale or above),
thus removing not only the cosmological problem involving the density of moduli,
but also the problem of vacuum destabilization for $H \gtrsim m_{3/2}$.  In the KL model, the constraint
$H \leq m_{3/2}$ does not apply, inflation is possible even in the simplest models of chaotic inflation where the Hubble constant may be extremely large, exceeding  $10^{13}$ GeV \cite{klor,Davis:2008fv}.  However, as we already noted,
uplifting in the KL model based only on anti-D3 branes in warped space leads to pure anomaly mediation, which is problematic from the point of view of phenomenology. Therefore, we were led to extended versions of the KL model  involving $F$-term uplifting.

In section 2, we have given two explicit examples of $F$-term uplifting
which generate scalar masses equal to the gravitino mass.  Relying once more on
strong moduli stabilization, the fields associated with uplifting do not
suffer from the Polonyi problem.
In the first example, we used a non-minimal form for the \K\ potential for the Polonyi-like
field $S$ \cite{Kallosh:2006dv}. The mass of $S$ was found to be much larger than the
gravitino mass which is sufficient for avoiding the Polonyi problem.
A second example based on the ISS mechanism \cite{iss} was shown to give qualitatively similar
results.

The main result of strong moduli stabilization with a decoupled uplift sector
is the generation of scalar masses equal to the gravitino mass, while
gaugino masses and supersymmetry breaking trilinear terms
are heavily suppressed by either the $F$-term of the modulus proportional to $D_\rho W \propto
m_{3/2}^2/m_\sigma \ll m_{3/2}$ or the vev of the Polonyi like field $\langle S \rangle \simeq
 \Lambda^2 \ll 1$  \cite{Kallosh:2006dv}. The resulting smallness of the $A$-terms
 necessitates a large gravitino mass to obtain a reasonably large Higgs mass as is now required.
 Thus we find a low energy spectrum built from very few parameters,
 reminiscent of split supersymmetry \cite{split}. Arguments for the decoupling of the uplift sector, which is important in obtaining
 these results, were discussed in Sections 2 and 3.

 However, given the theoretical framework for moduli stabilization described above,
 producing a viable low energy spectrum consistent with radiative electroweak symmetry
 breaking requires sensible supersymmetry breaking input parameters at some UV scale.
 Indeed, split supersymmetry models with universal scalar masses at the UV input scale
 put severe restrictions on the universal scalar mass, if these are input at the GUT scale
 as is commonly assumed in the CMSSM or mSUGRA models.
 While one could impose non-universality in the Higgs sector with non-minimal couplings
 between $S$ and the Higgs multiplets, we have assumed that this non-universality is not present
 in the simple models based on KL supergravity with O'KKLT uplifting.
 Instead, we have shown \cite{dmmo} that large scalar masses are in fact
 consistent with radiative electroweak symmetry breaking if the UV input scale
 for supersymmetry breaking is input above the GUT scale. In this case, the extra running
 between $M_{in}$ and $M_{GUT}$ generates sufficient non-universality
 for the low energy minimization conditions to be satisfied. Finally, to satisfy the
 boundary condition given by Eq.~(\ref{mubmu}), we include a Giudice-Masiero like contribution to
 the \K\ potential.

On the phenomenological side, the Higgs mass determination of $\simeq 125$ GeV may indicate significant splitting in the stop
sector (through a large $A$-term) or a high mass scale of supersymmetry.
Supergravity models with strong modulus stabilization choose the latter.
As we have seen the only ``light" supersymmetric particles in the theory are the
charged and neutral gauginos. Indeed the LEP lower limit on the chargino mass
coincides with a Higgs mass $\gtrsim 125$ GeV for $m_{3/2} \gtrsim 30$ TeV.
While gaugino masses and $A$-terms are generated predominantly through
anomaly mediation, the scalar sector of the theory resembles that of split supersymmetry.
However, unlike most split-SUSY models in the literature, the $\mu$-term derived here
through radiative EWSB  is of order the gravitino mass rather than
the gaugino mass.  Furthermore, the requirement of radiative EWSB
in split SUSY models is indicative of a supersymmetry breaking input scale above the GUT scale or of direct couplings of uplift fields
to the Higgs sector. The model described above, which followed the first option,  demonstrates that a UV completion of split SUSY models is
not only possible but is phenomenologically viable.

We have seen that despite the added complexity involved even in choosing minimal SU(5)
as a GUT, the low energy spectrum and in particular, the Higgs mass is relatively
insensitive to the choice of SU(5) couplings or the supersymmetry breaking input scale.
While there is some sensitivity to $\tan \beta$, the predominant sensitivity of the low energy
spectrum lies with $m_{3/2}$. The large value of $m_{3/2}$ required now guarantees not
only the absence of cosmological moduli problems (due to strong stabilization),
but also the absence of
any gravitino problem as well, as the massive gravitino decays well before nucleosynthesis.

While our starting point was the strong stabilization of moduli with appropriate $F$-term uplifting,
our resulting spectrum and phenomenological consequences resemble the
recent models of pure gravity mediation \cite{yana}.
 For a related analysis based on M-theory compactifications, see Ref. \cite{g2}.
Here, however, we have followed the full renormalization group evolution of
masses and couplings from the supersymmetry breaking input scale, $M_{in}$,
to the electroweak scale insuring consistent electroweak symmetry breaking.
 An important difference between our approach and that of \cite{g2} is our emphasis on the strongly stabilized versions of string theory/supergravity, which provides a simple solution of the problem of moduli destabilization during inflation, as well as of the cosmological moduli problem after inflation.
This approach led us directly to the phenomenological models described above.

There are several consequences of the derived sparticle spectrum.
Because squarks are heavy, gluino decay is suppressed.
However, given the ranges of gravitino masses considered here,
the gluino decay length still lies below the LHC resolution.
Because gaugino masses are determined by anomaly contributions,
the LSP is the wino and at the minimal value $m_{3/2} \sim 30$ TeV
the relic density of winos is far too small to make up the dark matter.
This may indicate that the gravitino mass is larger ($m_{3/2} \simeq 650$ TeV)
in which case, the wino mass is above 2 TeV and annihilations produce the
desired WMAP relic density (in this case, there is a one-to-one correlation between
the relic density and the gravitino mass); the reheat temperature after inflation is
sufficiently large so as to produce a sufficient density of gravitinos
which decay to winos and supply the correct relic density  (in this case, there is a one-to-one correlation between the relic density and the inflationary reheat temperature); or
simply,  something else makes up the dark matter (e.g., an axion).

In the case that the wino is the dark matter, we have seen that the elastic scattering
cross section on nucleons is far below the projected limit of a XENON 1T detector \cite{xenon}.
In contrast, the expected signal for gamma-rays from wino annihilation in the halo
is close to the current FERMI limit when the gravitino mass is close to its lower limit.

In summary, we have argued that models of strong moduli stabilization
lead to a very specific low energy spectrum and resulting phenomenology.
The spectrum resembles that of split supersymmetry with scalars in
the mass range 30 -- 1000 TeV.
Smaller masses result in an excessively light chargino, while higher masses
result both in an excessive wino relic density and excessive Higgs masses.
Thus from the assumptions of strong moduli stabilization and 
decoupling of the uplift sector, upon applying the LEP limit on 
chargino masses, we are immediately led to a  lower bound of roughly 125 GeV for 
the Higgs mass and a lower bound of roughly 1 TeV for the gluino mass.
Furthermore, the upper limit
on the gravitino mass from the limit on the wino relic density also corresponds to
roughly the current upper limit on the Higgs mass at the LHC.  Thus, there exists a narrow window
where all current experimental results are satisfied.
While direct detection of wino dark matter will be difficult in this model,
detecting gamma rays from wino annihilations
may indicate the realization of models of this type.

\section*{Acknowledgements}
We are very grateful to M. Arvanitaki, S. Dimopoulos, T. Gherghetta,  P. Graham, R. Kallosh,  J. March-Russell and X. Tata for many interesting discussions and suggestions.
The work of E.D. was supported in part by the ERC Advanced Investigator Grant no. 226371 ``Mass Hierarchy and Particle Physics at the TeV Scale'' (MassTeV) and by the contract PITN-GA-2009-237920. The work of E.D. and Y.M. was supported in part  by the French ANR TAPDMS ANR-09-JCJC-0146. The work by A.L. was supported by NSF grant PHY-0756174 at Stanford University.
The work of A.M. and K.A.O. is supported in part by DOE grant DE-FG02-94ER-40823 at the
University of Minnesota. The work of A.M. is also supported in part by DOE grant DE-FG02-04ER-41291 at the University of Hawaii.


\end{document}